\documentclass[preprint,12pt,authoryear,3p]{elsarticle}

\def\aj{\rm{AJ}}                   
\def\aap{\rm{A\&A}}                
\def\pasp{\rm{PASP}}               
\def\icarus{\rm{Icarus}}           

\usepackage{hyperref}

\journal{Astronomy and Computing}

\begin{document}

\begin{frontmatter}



\title{SSOXmatch: a Java pipeline to compute cross-matches of Solar System bodies in astronomical observations} 


\author[oan]{T. Alonso-Albi} 
\ead{talonsoalbi@gmail.com}

\affiliation[oan]{organization={Observatorio Astron\'omico Nacional},
            addressline={Alfonso XII, 3},
            postcode={28045},
            state={Madrid},
            country={Spain}}

\begin{abstract}

In this paper I will describe a new software package developed using the Java programming
language, aimed to compute the positions of any
Solar System body (among asteroids, comets, planets, and satellites) to help to perform cross-matches of
them in observations taken from earth- and
space-based observatories. The space telescopes supported are Hubble, James Webb, Euclid,
XMM-Newton, Spitzer, Herschel, Gaia, Kepler, Chandra, and TESS, although the flexibility of the
software allows to support any other mission without the need to change a single line of code.
The orbital elements can be selected among the asteroid database from the Lowell observatory (completed with
the cometpro database of comets maintained by the LTE), and the JPL database of minor bodies.

The software does not depend on external tools, and performs its own numerical integration of minor
bodies. The dynamical model implemented for the Solar System includes the gravity effects of all major
bodies, including the Earth, Moon, and Pluto as individual bodies, 16 perturbing asteroids as in
other tools, the General Relativity effects, the oblateness of the Sun, Earth, and Moon, and the
non-gravitational forces for both comets and asteroids. A complete set of
web services allow to compute the cross-matches (that are later to be confirmed, for instance by visual
inspection of the images) and also ephemerides of specific bodies. The code is highly optimized and follows the highest standards in terms of software
quality and documentation.
\end{abstract}



\begin{keyword}

numerical integration \sep N-body simulations \sep Solar System \sep cross-matching \sep ephemerides \sep web-service


\end{keyword}

\end{frontmatter}



\section{Introduction}
\label{sectionIntro} 

The observatories available on Earth and in the space provide a huge amount of data, that
can reach hundreds of gigabytes in each observing night, or even many terabytes in a near future for
the LSST survey of the Vera C. Rubin telescope \citep{jur17}. One of the products
is the serendipitious detection of minor bodies, that is delegated to automatic
pipelines following a number of steps, among them to clean the images from cosmic rays or
other artifacts, to recognize the presence of tracks of moving objects, and to
identify if they belong or not to known objects (see \citet{kru21, mah19}). When the astrometry and
photometry can be derived, these pipelines allow to catalog
new minor bodies, to improve the accuracy of the orbits of those already known, to
derive their size and rotation periods, and to clasify to which population the asteroids
belongs to (see \citet{kru21, rac22}). In these pipelines the accurate identification of the minor body is crucial to later
perform different scientific studies aimed to better understand the formation and
evolution of our Solar System, from the size distribution and chemical
properties of the asteroids, to more practical implications like the possible threat of
collision with Earth of Near-Earth Objects (NEOs).

Currently available pipelines present a number of issues that can limit the effectiveness of the
identification process. Most pipelines are executed against the image archives, not in real
time, since the process of identifying cross-matches can be slow when the potential
number of bodies to check is of several millions. Moreover, the numerical integration for
such a huge amount of bodies is also too slow to be executed in real time, which forces
the bodies to be pre-integrated during decades in steps of time, a
process that can be inefficient when this pre-integration is performed in each survey individually.
Another issue is that every observation covers a particular region in the sky, and is spread
over a time interval, that can be minutes, hours, or even more. But most
pipelines provides the cross-matches for a given radius around a point in
the sky (a cone-search, as defined by the \href{https://www.ivoa.net/documents/ConeSearch/20200828/WD-ConeSearch-1.1-20200828.html}{IVOA specification}),
and for a fixed instant, not all those that took place over a given
time window and specific sky region. This is the consequence of using software libraries that
were designed to provide ephemerides, and not
specifically written to identify efficiently which minor bodies were present in a given sky
area and time window, from a list of millions of bodies and observations, that is continuously growing.
Another consequence is that most of these services are somewhat limited, providing only the bodies present
in a given field and instant. 

From the limited number of tools available to identify which minor bodies were present
in the field of view of a given observation, we can find SkyBoT \citep{ber06} and the
\href{https://ssd-api.jpl.nasa.gov/doc/sb\_ident.html}{JPL small body identification tool},
which is part of the \href{https://ssd-api.jpl.nasa.gov/}{JPL SSD/CNEOS API Service}.
Most other pipelines use one or the other service
for the crucial step of computing the positions of minor bodies \citep{kru21, mah19}. The tool by JPL uses
the Horizons ephemerides server, that provides high-accurate ephemerides of minor bodies, but
the service is frequently slow and may throw timeout errors for queries some years in the past. In
addition, this service does not support space telescopes, unless the position of the space telescope
is provided by the user.
The \href{https://ssp.imcce.fr/webservices/skybot/}{SkyBoT tool} performs fine, but the dynamical model
used in the numerical integration suffers from issues that may limit its applicability to NEOs (see Sect.
\ref{sectionPhysModelCompareSkyBoT} and \ref{sectionWebComparison}), and the options supported for
space surveys are also limited.

Other pipelines are based on the software behind the previous ones. This is the case of the pipeline by
\citet{rac22}, in which the Eproc software\citep{ber98} is used to perform the numerical integration
of minor bodies, as in the SkyBoT tool. This work is based on \citet{rac19}, an effort
to integrate a mechanism to search for Solar System bodies in the astronomical observations shown in the
ESASky portal \citep{gio18}. To compute the cross-matches of minor bodies in the observations of different
surveys, the Eproc software is called multiple times for each asteroid until its presence in
the field of view of a given observation is confirmed or discarded. Despite the numerical integration in
Eproc is fast, the performance of the cross-matching process of this pipeline is limited, even when executed
in a grid infrastructure with around a hundred of threads. Due to the increment expected in the number of
minor bodies discovered and observations available in the next years, it was anticipated that it could
become too slow to be practical.

The conclusion was that a new generation of software tools designed specifically to compute cross-matches
was required to overcome these practical and technical limitations, to be ready in the near future for a
massive amount of observations and minor bodies, many of them NEOs. This is the origin of the SSOXmatch
package described in this paper,
%
a new development aimed to perform an accurate, flexible, and efficient
numerical integration of asteroids and comets. The paper is organized as follows:
in Sect. \ref{sectionComputations} I present a detailed description of the calculations
performed by the software, including as subsections the dynamical model, the numerical
integration schema, the particularities of the orbital elements in the JPL and Lowell databases,
and how the positions of the natural satellites
were computed. A number of problems were identified and
solved during the development. In Sect. \ref{sectionValidation} a comparison of SSOXmatch with other
integrators is presented, for both long- and short-term test cases, to
justify that the integrator is suitable for the problem of computing cross-matches of minor bodies.
The collision of comet Shoemaker-Levy 9 with Jupiter is presented as an additional consistency
test with Horizons. Additional subsections show a general comparison of the orbital elements between the
JPL and the Lowell databases, as well as a comparison of the dynamical model implemented with respect to the one
used by \citet{ber06}. In Sect. \ref{sectionXM} the cross-matching process is described,
detailing the filtering of the observations to discard those with inconsistent data, the
optimizations implemented to discard as soon as possible bodies that
will never appear in a given field, the direct approach implemented to identify a
cross-match, and the propagation of positional uncertainties. Sect. \ref{sectionXmResults} presents
a general study of the results of the large scale computations of cross-matches for the
surveys of some space-based observatories. Sect. \ref{sectionWeb} describes the web
services implemented, that covers the functionalities of others from
LTE or JPL, while providing additional functions. A comparison of the services is
provided in Sect. \ref{sectionWebComparison}. In Sect. \ref{sectionConclusions} I summarize
the main results of this work.

\section{Computations}
\label{sectionComputations}

The SSOXmatch package was written in the Java programming language. It contains 40 000
lines of code spread in 76 files, with an additional 12 000 lines and 45 files of tests and utilities.
The first natural question would be why this language was chosen for such a large project instead
of Fortran, which is the traditional language used in this field. There are so many reasons
that a specific paper could be written about this. An interesting lecture could be the Sections 4
and 5 of the paper by \citet{che21}. The key point is not if Java programs are fast, which is the case,
but more importantly the tools a given language provides for extremely important tasks related
to the development and maintenance of the source code, like the Integrated Development Environments (IDEs, to
avoid the risks of writting in text editors),
which are usually more advanced in Java than in other languages; the object-oriented approach,
which is an extremely useful and life-saving programming paradigm to organize and manage huge amount of
data efficiently using objects; the javadoc documentation within the code that allows to directly read your
own comments and help information from other pieces of the code; the profiling tools like VisualVM,
useful to monitorize the programs and to identify performance or memory
bottlenecks; and the `compile once, run anywhere' philosophy behind Java, that allows
to run programs properly written in any platform and version of Java, without
ever re-compiling the code. And there are still many additional features in the IDEs aimed to help
to maintain the quality and readability of the code, essentials for large projects. Java
programs can be easily integrated in other projects and server tools, for instance for the Virtual
Observatory.


\subsection{Presentation}
\label{sectionComputationsIntro}

The entire SSOXmatch software is contained in a single, 2.0 GB jar file, which is a compressed
zip file internally. Inside it there is a structure of files and folders including the orbital elements
of asteroids and comets (for the two databases available), the complete list of observations for the space surveys supported, the
kernels and bindings, the web service files, and the documentation. Everything can be run
or accessed from the command-line.

SSOXmatch uses the Java bindings for the SPICE toolkit\citep{acton96,acton18} to
compute the positions of planets and spacecrafts from bsp kernel files. Although these bindings
are officially in alpha status, they have proven to work very well since years, and the documentation
and examples are very useful to get started effortless. This toolkit can be
distributed inside the software, with the correct native library for a given platform read during
the startup, so that multiple platforms can be supported and no re-compilation of code or particular
packetization is needed
at all. The philosophy of Java `compile once, run everywhere' still applies. The supported official
platforms are 64-bit Linux, Mac, and Windows systems, and the Solaris platform. In this package
the support for Solaris was not included, but support for the ARM-64 architecture was added
to Linux and Mac systems, by performing a compilation of the SPICE toolkit for these systems and
architecture, a process that was smooth.

With SPICE any of the JPL ephemerides integrations can be used by reading the
corresponding kernel file. There are multiple integrations supported,
from DE200 to DE440, and for each of them the constants used in the original integration are
read from the JPL header text files included in the jar file (that can be found in the web along
with the JPL integration files in plain-text
format), to adapt all the internal constants of the integration to the original JPL integration.
This affects the masses of all major bodies and perturbers, the oblateness coefficients, the Astronomical Unit (AU),
the obliquity of the ecliptic, the Earth/Moon mass ratio, among other values. Before DE430 the au
was a free parameter that was fitted in each ephemerides according to the mass of the Sun derived and
the Gauss constant k = 0.01720209895, from the expression GM$_{\odot}$ = k$^2$. Starting with
DE430 \citep{fol14} the au was redefined and fixed to 149 597 870.7 km, according to the resolutions
adopted by the International Astronomical Union (IAU), and the Gauss constant was left as a free parameter.
As a result, a new parameter appear in the
header of the DE430 and later integrations, representing the rate of change with time of the solar mass (GMSDOT). There is
also a J2EDOT to account for the rate of variation of the oblateness of the Earth. These
details are naturally and automatically handled by the software. For DE440, the masses of the
Kuiper Belt Objects (KBOs) and the Kuiper Belt ring model added to the dynamical model of
the Solar System \citep{par21} are also supported.

The numerical integration schema implemented supports the treatment of collisions, a feature
that is not frequent in other works. Combined with the orbital elements from the JPL
database, it is possible to compute cross-matches of bodies that collided with any planet or asteroid,
like the collision with Jupiter of each of the fragments of the comet Shoemaker-Levy 9, that was intensively observed by
the Hubble Space Telescope in 1994, or the collisions of around a dozen NEOs with Earth little after
they were discovered, like 2008 TC3 or 2024 BX1. During the pre-integration phase SSOXmatch
will compute a list of bodies that suffered close encounters with other bodies with mass,
including perturbing asteroids. These bodies can be considered as critical, in the sense that
a rigorous treatment of the interactions is required to propagate accurately their positions
beyond the time the close encounter took place. The numerical integration in SSOXmatch has been
implemented carefully,
but in some cases there is a little discrepancy with Horizons that can get amplified with time.
One of such critical examples is 2020 CD3, a body that behaved almost like a temporal moon of the
Earth. When integrating back with elements referred to the current epoch, the consistency with Horizons
starts to decline after the third close encounter with Earth (for this particular body Horizons considers
the nonzonal harmonics of the gravitational field of the Earth, caused by azimuthal asymmetries,
see \citet{hol23}, something not implemented in SSOXmatch). To solve this situation the
kernel file for this and another one thousand bodies were downloaded to cover the time interval in
which these close encounters took place. For consistency these kernels are only used
when JPL elements are selected, and, outside the critical time window, the pre-integration
continues with the common numerical integration process implemented. As mentioned, this takes place during
the pre-integration, a process that is extended into the past until year 1992 (for the first
observation in the Hubble archive, which is also the oldest one in this software), to write the barycentric position and velocity vectors for each body
in intervals of 73.05 days (five times per Julian year). This pre-integration is unique for all surveys,
generating currently around 27 GB of data, an amount of space that can be stored in memory in a
reasonably powerful machine.

\begin{figure}[t]
\centering
\includegraphics[width=1.0\textwidth]{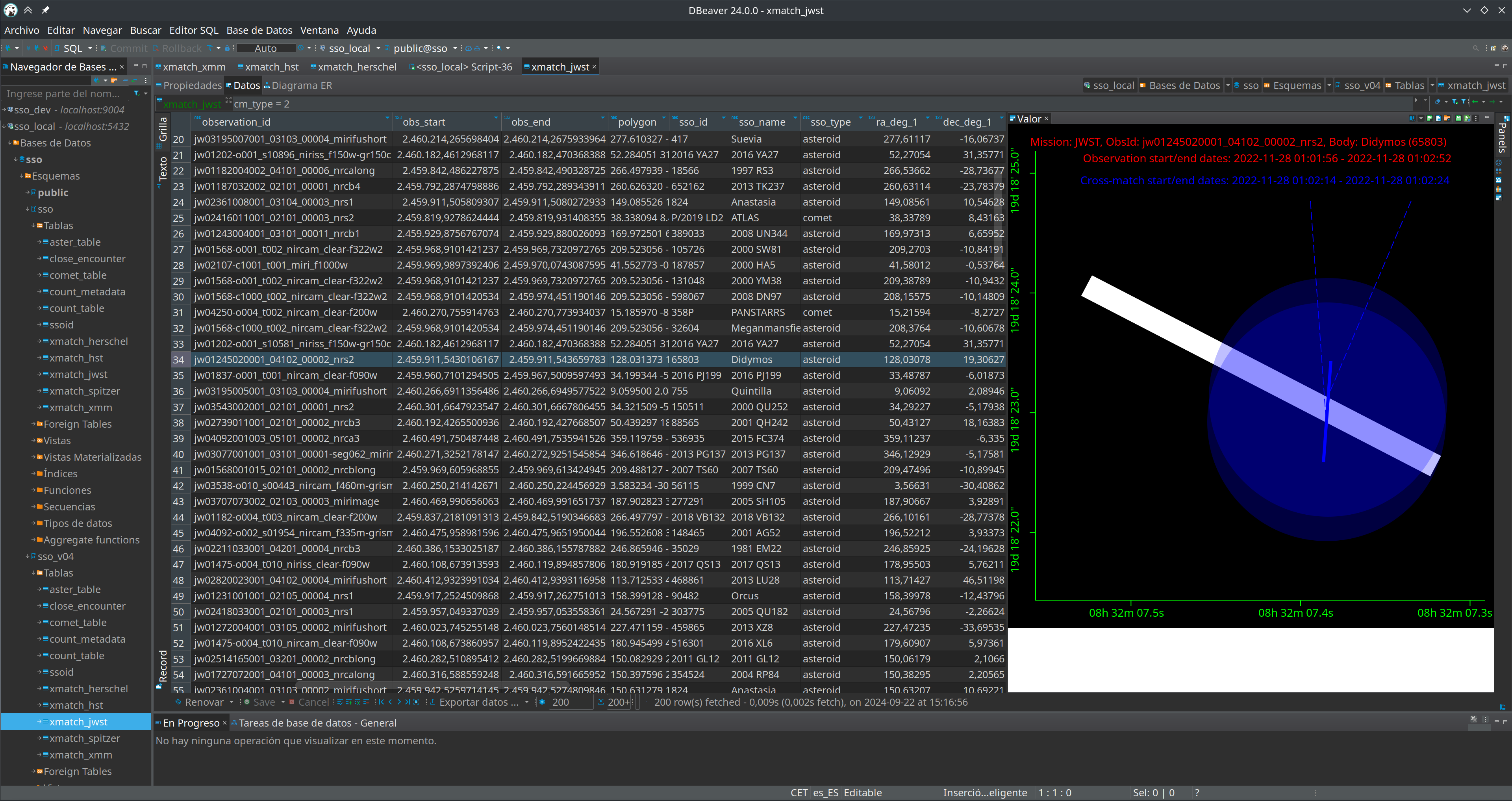}
\caption{The screenshot shows the DBeaver database management tool being
used to explore the cross-matches computed for JWST and ingested in a postgres database,
and to visualize a sketch of one of them. The sketches are also ingested for every cross-match.
Some of the database columns generated by the software are visible.}
\label{figDBeaver}
\end{figure}

In SSOXmatch most files can be automatically updated by the user,
including the kernels for the updated trajectory of the spacecrafts, the list of observations,
the orbital elements, among other files that may need an update eventually. The new files are
copied to the main folder of the program creating a structure of files and folders identical
to the one inside the jar file. When an external file is present and it is newer than the one
inside the jar file, the external one is used instead. Once the elements
are updated the pre-integration must be repeated, a process that can take around a week in a
modern computer if all the quality flags are in the maximum level (the web service can perform
incremental updates automatically, as it is described in Sect. \ref{sectionWeb}). The
update of these files or the observation lists are performed with queries to external web
services or databases, and the commands needed are included in the properties
file. This file is key since it provides the values selected by the user for all the parameters
or options available, to control 
how the numerical integration or the computation of cross-matches are
performed. For instance, the cross-matches can be written to text files or to a postgres
database, and some of the columns generated with the information are visible in Fig. \ref{figDBeaver}.
The output optionally includes charts showing the footprint of the observation and the
trajectory of the cross-matching body, in SVG or JPG formats (see Fig. \ref{figDBeaver} and
\ref{figXmHST}). The sketch can optionally show the
background image of the sky from the Digital Sky Survey (DSS). This is accomplished with two
dependencies included in the jar file. By editing the properties file it is also possible to
support any other mission.

\subsection{Dynamical model}
\label{sectionPhysModel}

The dynamical model implemented considers the gravity effects of all planets, including the Earth,
the Moon, and Pluto as individual bodies. To account for the particular interactions of the Earth-Moon
system an implementation of the simple Earth-tides model used in the DE102 and DE200 numerical
integrations \citep{new83} can be enabled. The number of perturbing asteroids is arbitrary, between the
first one (Ceres) and the entire set of 343 perturbers in the latest JPL integrations. It is also
possible to include the 30 KBOs present in DE440 \citep{par21}. The General Relativity (GR) is
implemented in a flexible way, so that different formulations can be applied, mainly the simplified
formulation by \citet{dam85}, and the complete Einstein-Infeld-Hoffman model, also called PPN
formulation, described in the DE430 and DE440 technical documents. The
calculation of oblateness is supported for all major bodies, but in practice in most cases
only the J2 term of the Earth is enabled, as in Horizons, and the J2-J4 terms of the Sun, Jupiter,
or other bodies are not. For instance, the trajectory of comet Shoemaker-Levy 9 will match the
one reported by Horizons, but only when the oblateness of Jupiter is not considered.

It is important to note that some integration options need to be tuned carefully. Both
in the JPL and Lowell databases the elements
are referred to the current epoch (except for comets, and some asteroids in the JPL database).
When a given NEO suffered a close encounter ten years in the past,
an accurate propagation during these ten years is a must to reproduce the correct positions
when it was close to the Earth, and observed (with potential cross-matches). But how can we define
a rigorous propagation?
The answer depends on how the orbital elements were originally fitted to observations, and
specially how they were propagated into the future to list their elements referred to the
current epoch. For the integration of minor bodies SSOXmatch applies the same criteria used in the
Horizons ephemerides server: the oblateness is limited to the J2 term of the Earth, and the GR
activated is the PPN formulation restricted to the Sun. But in general, it is mandatory
to understand how the particular orbital elements were fitted and propagated, otherwise it is
not possible to implement a similar model to propagate them back years into the past and to
reproduce the observations that were used to fit them. For instance, some orbit fitting tools like
FindOrb \citep{gra22}, include additional physical effects not considered in the calculations for the
JPL and Lowell databases.

In Horizons, the elements used are referred in some cases to a date
around that of the observations, and not propagated to current times. These elements can be obtained by downloading the DASTCOM5
database \citep{gio20}, but a specific reader for this database would need to be implemented. Having elements referred
to the date of the observations is desirable, since most NEOs will not be observable again until many
years later, and the positional uncertainty will become high, making the elements somewhat meaningless.
Despite of this, it is remarkable the high degree of consistency of the
elements in the JPL and Lowell databases, that can be used to accurately reproduce the astrometry of
the observations (see Sect. \ref{sectionPhysModelJplLowellElem}).

In SSOXmatch there are some approximations considered in the numerical integration. The computation of
the oblateness does not consider the precession and nutation of the Earth axis, and the axes of the Sun and
the Moon are also considered to be static at J2000 epoch. The consequences
were tested to be negligible, and the omission practical, considering the slowdown these effects introduce
in the computations.
The JPL integration computes the position of the Moon relative to the Earth, to improve the accuracy
of the integration. This is hardly needed in SSOXmatch, since in practice the positions after
each iteration of the integration are replaced by the vectors from the JPL ephemerides kernel.

\subsection{Numerical integration process}
\label{sectionIntegrationSchema}

The integration process uses a generic implementation of an explicit Runge-Kutta (RK), that can be
used with more than ten solvers of orders between 5 and 12. Each of them includes an embedded
schema of lower order, between 4 and 10, that is used to estimate the integration error and to
adapt the integration step. Among the schemas implemented we can find the Bogacki-Shampine 5/4 \citep{bog96},
Dormand-Prince 5/4 and 8/7 \citep{dor80, pri81}, Tsitouras 5/4 and 6/5 methods \citep{tsi11, pap96},
Verner 6/5, 7/6, and 9/8 \citep{ver93}, and the Ono 12/10 schema \citep{ono06}. The default schema selected
for the numerical integration is the Bogacki-Shampine 5/4 (BS54) integrator, which is slower but slightly
more accurate than other solvers of the same order, although other 5/4 integrators like the
Dormand-Prince one are also suitable. For this work high-order RK did not show any particular
advantage over lower order ones, the main difference is they are slower.

In the literature implicit RK solvers are considered more adequate than explicit ones, because
their region of stability is greater. However, these solver are also considerably slower, frequently
more than an order of magnitude (\citet{lof13}). The advantage of implicit schemas are evident in stiff
problems, where the forces involved present a high gradient in space and time. Explicit RK will
use extremely little time steps in this situation to maintain stability, which would make them slower
than implicit ones, and the accuracy they can get would get limited due to the round-off errors introduced
by the high number of steps. In this particular work,
this can happen in close encounters between bodies in which at least one has mass,
an event that is exceptional. For instance, if we set as constraints to consider an encounter as a
close encounter the distance limit of 0.001 au, or 40 times the radius of the massive body, the sixteen
perturbing asteroids usually considered to have mass will produce just above 60 close encounter with other,
little minor bodies between 1992 and 2030. Jupiter and Saturn have around 50 close encounters, most of them
with comets, while Mercury, Venus and Mars just a few. The other 900 are with the Earth and the Moon,
produced by around 800 distinct bodies. A fraction of these bodies may require a sophisticated solver to
accurately propagate their positions beyond the close encounters, but an explicit RK will still be a good
option outside that specific window. In this work the goal is to deal with millions of bodies and
observations, and to be ready for a future in which many more millions will come. Since the performance is
as important as the accuracy, an explicit RK schema was selected.
To keep the accuracy in close encounters, the kernels for the particular bodies, around 1000, presenting
close encounters were downloaded from Horizons for the specific time windows computed for the close
encounters, and used instead of integrating them. Another option would be to download and to try to keep
updated the kernels of all the bodies, for a period of around 40 years. This would take many terabytes of data,
which could still be technically possible, but it is unclear if this would be the case in the future. In
addition, it is also unclear if SPICE could handle such a huge amount of bodies.

The integration process is applied to all bodies including planets. After each integration step,
or in intervals of some steps, it is possible to call the SPICE toolkit to replace the positions
of the planets by those returned by the JPL ephemerides. The SPICE toolkit is not a thread-safe
library, so each calculation must be called
while no other thread is interacting with SPICE. When the calculation is performed with
hundreds of threads, this limitation can almost freeze the program, despite how fast
SPICE is. The problem was minimized with the implementation of a cache system, that avoids
repeating the calls to SPICE with identical inputs from different threads. Another option is
to not replace after every iteration the positions by those coming from the JPL ephemerides, but
after several integration steps, like
ten or one hundred, although the default process followed is to replace them after each iteration.
This ensures consistency in the sense
that the barycenter of the Solar System is kept at the same location when the number of perturbers
does not match those used in the JPL ephemerides.


When all the contributions to the acceleration of the body are computed, the result are a set of
numbers that need to be added up, with a wide range of orders of magnitude. To perform an accurate
sum may become a problem with the typical accuracy of double-precision values (16 significative digits).
Consider this example: you may compute the gravity force of the Sun for a given asteroid, or
the Earth for a NEO, obtaining an intense force, and later try to sum to this force the effects
of a large number of low-mass perturbing asteroids that are located farther. In case the intense
force is equal to 1.0, and you want to add later another term of 1.0$\times$$10^{-17}$, you will end with 1.0 after
the sum. In case you sum one million contributions of this amplitude, you will still have 1.0
as result, so all these little contributions are not added to the total force. To correctly
deal with this the sum should be performed first to the low-mass perturbing bodies, or in
general all these contributions added to a list and summed in crescent order of the absolute
values. In
that case, one million contributions of +1.0$\times$$10^{-17}$ will become 1.0$\times$$10^{-11}$,
that can be added to 1.0 later with 16 digits of precision. In addition to this, there are
strategies to reduce the round-off errors produced when the number of elements in a sum is large.
The most famous algorithm is the Kahan compensanted summation algorithm \citep{kah65}, later improved by
\citet{kle06} by adding a second-order iteration. SSOXmatch supports to sum the contributions in crescent
order, as well as to apply the Kahan or Klein algorithms, which also require to sort the values.
The ideal situation would be to perform the computations
in extended precision arithmetics, using for instance 32 significative digits. This is partially supported
in SSOXmatch, in the computation of the gravity contributions, but it is of little practical
use since it is extremely slow. Detailed tests show that for an integration of several decades
the summation in crescent order can provide some additional significative digits of precision,
although this obviously depends on the particular body and time period.
In general, this effect is not important, since it only improves the output vectors a
few km, but a detailed study may show some specific relevant cases.
In Horizons, the calculations are performed natively with
18 significative digits of precision due to the special machines used, so in comparison this
problem is of little or no importance. In many numerical integration packages, specially
those based on old code, this problem is not considered, while in modern ones it is recognized. For
instance, in ASSIST \citep{hol23}
the gravity effects are computed starting from the less massive body, adding the effects of the
Sun at the end. In addition, the IAS15 integrator used supports the Kahan compensated summation, so this approach
can accurately reproduce close encounters with NEOs.

As mentioned previously, all the detailed options for the numerical integration, including
the integration schema, perturbers, treatment of GR and oblateness, how to
sum the force contributions, among others, are selectable with the options present in the
input properties file. This flexibility in the implementation allows to construct a customized
Solar System. In addition, the calculations were optimized as much as possible, for
instance by computing the minimum number of square roots for the distance between the bodies.

\subsection{Orbital elements}
\label{sectionElements}

In SSOXmatch there are two asteroid databases available: the JPL small body database, and the Lowell database (also
known as astorb database). The MPCORB database from the Minor Planet Center (MPC) was also considered, but
discarded in favor of the JPL one. Among the reasons the need of similar elements to compare the results
with Horizons, including the non-gravitational parameters (difficult to find in other databases), to properly check the dynamical model, and the
number of decimal places in the values of the elements. In the MPC database there are seven decimal places
for the semimajor axis, which means an initial position uncertainty close to 10 km. The Lowell database provides
an additional digit, while the JPL one all the digits to keep the consistency and allow comparing the output
in double precision computations.

The JPL minor body database can be used to download the elements of all known asteroids and comets,
including those that collided with Earth or other bodies. They also provide the updated
non-gravitational forces parameters of comets and around 500 asteroids (most of them NEOs), that
are mandatory data to accurately reconstruct the orbit trajectories. For comets these parameters
account for the trajectory change produced
by the ejection of material from the tail of the comet, following the model by \citet{mar73}, while for
asteroids they allow to model the effects of the solar radiation pressure and the
Yarkovsky effect \citep{far13}.

The Lowell database of asteroids \citep{mos22} lists all bodies that exist today, so it does not
list bodies that collided with Earth. Some relevant bodies like the interestellar asteroid Oumuamua,
that was first cataloged as comet and later classified as asteroid, are not present either. In
\citet{mos22} some details about the Lowell database
are provided, explaining that the non-gravitational parameters of asteroids have been
considered and taken from the JPL database of minor bodies. The problem is that these coefficients
frequently change, and there is no way to know which values were used, since their values are not
included in the database of elements or the asteroid information pages. There are cases like the
asteroid Didymos, for which the values have changed over the time until a point in which this
body currently has no non-gravitational parameters assigned in the JPL database. These changes are consistently
applied to the ephemerides server Horizons, but it is unclear if
they are needed or not in the Lowell database, difficulting the accurate reproduction of the correct
trajectories. Respect comets, the LTE database cometpro
\citep{roc96} does not include the Shoemaker-Levy 9, and many other comets that are included in the
JPL database, showing some lack of completeness in the database. The non-gravitational parameters in
this database does not include the time-delay parameter, which can also limit the
modelization of the trajectory of some comets.

During the development and testing of SSOXmatch a problem was found in the non-gravitational parameters
reported in the JPL minor body database: they are generally slightly different from those
reported and used by Horizons. Despite the difference is well below the uncertainty in these parameters,
this problem prevents an accurate reproduction of the results of Horizons for tests involving a number
of critical bodies. Moreover, the specific model these parameters are referred
to (parameters named ALN, NK, NM, NN, and R0 in Horizons\footnote{These parameters are used to compute the
main model force parameter g = ALN q$^{-NM}$ $\times$ (1 + q$^{NN}$)$^{-NK}$, with q = r / R0. R0 is the
sublimation radius, ALN a normalizing factor, NM the primary radial scaling law exponent, and NN and NK are
distance exponents. This value is later used with the non-gravitational parameters to compute the non-gravitational
force in the radial, orbit-transverse, and orbit-normal directions.}) are not provided in the small body database, and there
are bodies with particular models, for instance Toutatis or 2008 GO98.
As a workaround to this problem, the non-gravitational parameters downloaded from the JPL database are later
replaced and completed automatically by those used in Horizons, for both the non-gravitational parameters and
the model. This step is also needed for comets.

The trajectories of the minor bodies are calculated by solving the Kepler equation using a Newton-Raphson iteration for
elliptic (eccentricity below unity) and hyperbolic (eccentricity above unity) orbits. All bodies
available were tested carefully to ensure that the iteration always converge, something that in some
critical cases required to repeat the iteration starting with a different, more adequate guess
for the eccentric anomaly (see \citet{dan83, dan87}). In case there is no convergency in any future situation, an error would
be thrown, which ensures a wrong calculation is not possible as an output from the software.
For perfectly parabolic orbits the equation is solved analytically \citep{mei85}. The clever
analytical solution by \citet{mik87} was also studied, but discarded since it does not provide
the same result as the Newton-Raphson iteration up to the last significant digit, specially when the
eccentric anomaly is close to zero, or the eccentricity high. This led to slightly wrong initial integration
vectors for the positions and velocities, as it was checked against the Horizons server, with the subsequent
amplification of the deviation in the numerical integration.

\subsection{Natural satellites}
\label{sectionMoons}

To compute the positions of natural satellites there are two main set of algorithms implemented. One
is an approximate treatment based on precessing ellipses, using elements provided by the JPL
in their web pages. This approach is simple, generic for all bodies, and very fast. The objects
supported are the two main satellites of Mars, the four main satellites of Jupiter, the six
main satellites of Saturn (excluding Hyperion, Iapetus, and Phoebe), five satellites for Uranus,
and Triton for Neptune. The maximum errors observed are in Jupiter, around 3" for Ganimede
and Callisto, but the median of these errors are only slightly above the arcsecond. The rest of bodies
are accurate to 1" at least. The second approach is to use the implemented analytical
theories that provide better accuracy, among them the \citet{lai07} theory for the Martian
satellites, the L1 theory for the satellites of Jupiter \citep{lai04a,lai04b}, TASS 1.7 for
the satellites of Saturn \citep{vie95,dur97},
and GUST86 for Uranian satellites \citep{las87}. The support for other satellites is possible by means of numerical
integration. This is the case of Phoebe, the external, highly perturbed moon of Saturn, and
Nereid in Neptune. It is interesting to mention that the numerical integration of these two
satellites is performed on-the-fly with the same code used to compute the integration of the rest of bodies, so
there is no need of external files with Chebyshev polynomials, or any other kind of approach,
to naturally support these satellites. SSOXmatch does not support irregular satellites or satellites of asteroids,
but support for them could be added in the future.

The planetocentric positions of natural satellites are simply added to the barycenter of the
planets, and later corrected for light-time. The barycenters of the planets with satellites
are obtained using DE440 ephemerides. There is a little offset between the physical
center of the external planets and their barycenters, that is ignored in the calculations of the cross-matches
of planets. In addition, the gravitational effects of the satellites are not considered.
There is no practical effect of these approximations in the output of the software.

\subsection{Reference frame and topocentric corrections}
\label{sectionRefFrame}

The coordinates computed by SSOXmatch are astrometric J2000. These coordinates are compared later with
the observations, that are also astrometric and referred to the J2000 epoch. Depending on the survey, these coordinates
may be expressed in the ICRS or the FK5 reference systems, but no explicit frame transformations are
performed in SSOXmatch to accurate align the coordinates in the different surveys. The offset between these
systems is around 0.02", which is negligible in comparison with the field of view of the observations.

For the computations of ephemerides, the purpose during the development was to reproduce as close as
possible the output from Horizons, when using the set of orbital elements from the JPL minor body
database. The algorithms were selected for this purpose, and implemented to maximize the performance
without losing accuracy in the level of a few 0.01". The topocentric position of the bodies
requires the computation of the local sidereal time. For this calculation, the nutation is
implemented with the main terms of the IAU2000A nutation model, only those with an amplitude greater
than 0.01". The difference with respect to the IAU1980 model used in Horizons is in the level of few
milliarcseconds. The UT1-UTC difference is computed by means of an EOP file that is updated
from the IERS website automatically. Planetary aberration is corrected to first order in general,
but a second order correction is applied for bodies closer than 0.5 au to the telescope. Light-time
is also corrected with respect to the geocenter for Earth-based observatories. Planetary magnitudes are computed following
\citet{mal18}, in which the classic formulae is extended to compute the magnitudes of the planets for a
wider range of phase angles, like those present in space surveys. This is also the reference used by Horizons.
The topocentric correction also requires a rotation for the vector representing the geocentric
position of the observer, from the true equinox of date to the mean equinox of J2000. The precession correction
in this rotation is computed using the method by \citet{lie79}, like in Horizons.

\section{Validation of the numerical integration}
\label{sectionValidation}


\subsection{Long-term numerical integration test}
\label{sectionTestLongTerm}

This test uses as the only input the position and velocity vectors for the planets and all 343
perturbers included in the JPL DE432 ephemerides, returning the output vectors from SSOXmatch in
intervals of five years, and comparing them back with DE432. This version was selected instead of
DE440, which is the default integration used to initialize the bodies for computing cross-matches,
because DE440 includes a model for the Kuiper belt, with 30 additional perturbers and a set of rings,
that, although they have been implemented in SSOXmatch, this implementation is still not consistent
enough with DE440 to show the overall behavior of the implemented integrator.

The options selected in the properties file to mimic the results from the DE432 includes all the
physical effects: oblateness of the Sun, Earth, and Moon (more specifically, the J2, J3, and J4 terms, as
explained in \citet{fol14}, but considering the orientation of the axes are fixed at J2000 epoch), and
the complete PPN formulation (but including only the three most massive asteroids Ceres, Pallas,
and Vesta). The BS54 integrator was
selected, and the Kahan compensated summation algorithm enabled. The barycenter of the system was
reset after each integration interval to maintain the same position and velocity (almost zero) found
when the integration started.

\begin{figure}[h]
\centering
\includegraphics[width=1\textwidth]{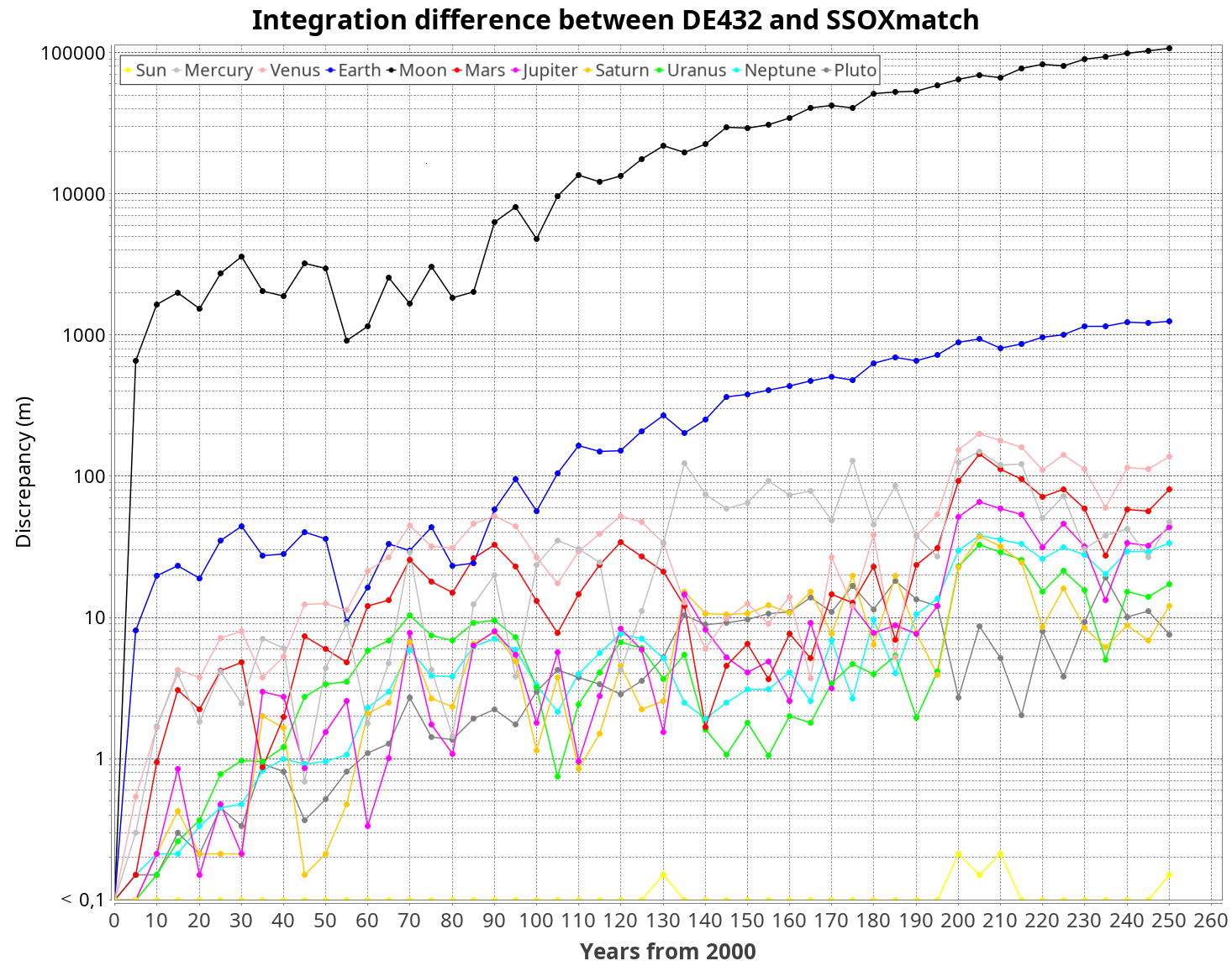}
\caption{Comparison of the numerical integration in SSOXmatch with DE432 during 250 years, starting
from year 2000.}
\label{figValidation1}
\end{figure}

The results of the comparison are shown in Fig. \ref{figValidation1}. The agreement is
generally better than 100 m after 250 years, except for the Earth-Moon system. This result was
expected since in the implemented integrator the vectors for the Earth and the Moon are treated like
those for other bodies. It would be better to use the barycenter of the Earth-Moon system to
integrate only this body in the RK. In addition, the DE200 Earth-tides model was not enabled, and, despite of
being very basic, this model improves the results by a factor five, to a final discrepancy of
30 km for the Moon and only 200 m for the Earth. The precession and nutation
of the Earth were not considered, since it was found to provide a minimum improvement with a considerable
penalty in terms of integration performance. As mentioned above, the GR option selected was not
extended to all asteroids, since that would be very slow. All these approximations are affecting the results,
considering we are in the level of 10-100 meters after more than two centuries. It is noticeable
that the inner bodies deviate more than the outer ones, maybe because the integration step is the
same for all bodies and the accuracy of the RK is lower for bodies with higher angular
velocities. But considering that this integration mode is never used in practice, because the
default computation mode for cross-matches is to replace the planetary vectors by the JPL integration
after each integration step (so we get the effects of all perturbers without the need to compute
them individually), the results obtained show that the numerical integration is suitable for the
purpose of the present work, which is to calculate high-accurate, short-term integrations of minor
bodies. The quality of the integration and other
operations performed in the package is also supported by more than one hundred JUnit tests.

\subsection{Short-term numerical integration test: Apophis}
\label{sectionTestApophis}

The second test was to repeat the same test of the close encounter of Apophis with Earth described
by \citet{hol23}. In their software ASSIST they obtained a discrepancy of around 500 m (with respect to the integrator
used at JPL for small bodies) after 1000 days of integration, starting from January 1, 2029, and continuing with the
integration after the close encounter in April, 13. The final discrepancies are mainly
the result of the little discrepancies in the velocities produced during the close
encounter.

The same test was conducted for the numerical integration implemented in SSOXmatch, but in this
case comparing the results with Horizons. It is important to note that in this work it was not possible to
access the small body integrator or perform deep specific tests with Horizons, so the comparisons are only
indirect by means of numerical tests. For the
computations the BS54 explicit RK was used, starting from the same vector
returned by Horizons, and comparing the positions in intervals of 5 days. In this case,
the integration options should be adapted to allow a comparison with Horizons: the oblateness is
limited to the J2 term of the Earth, the GR is limited to the Sun in the outer loop of the
formula (see \citet{par21}), while no asteroids are considered in the inner loops (like in ASSIST), and the number
of perturbers is limited to 16. All these options should be carefully selected, otherwise the
comparison is not correct. It is noticeable the practice in Horizons to discard the oblateness
of non-rigid bodies, like the Sun or the external planets.

\begin{figure}[h]
\centering
\includegraphics[width=0.49\textwidth]{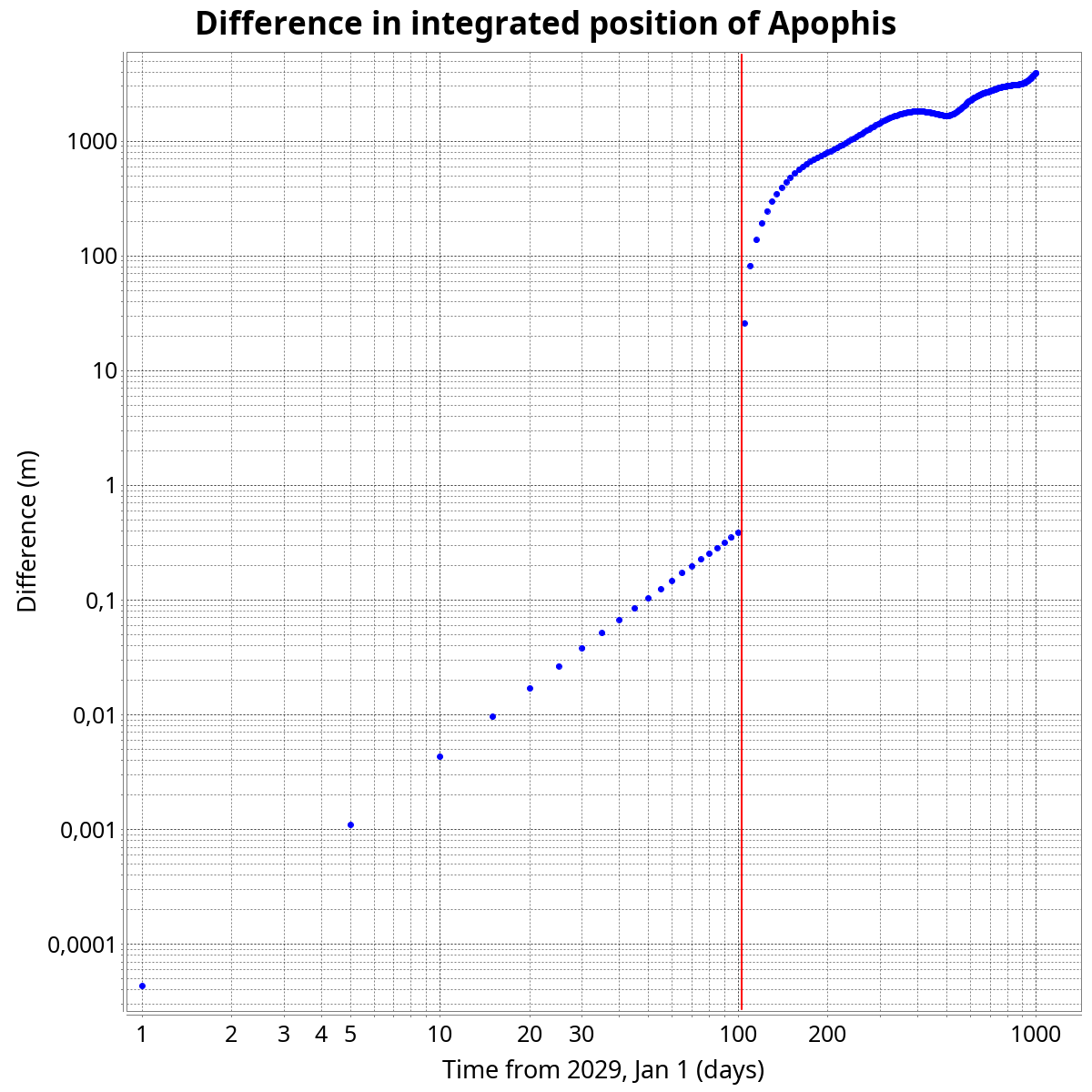}
\caption{Discrepancies between SSOXmatch and Horizons for the close encounter between Apophis
and the Earth (1000-days integration starting from January 1, 2029). A vertical red line represents
the close encounter time on April 13, 2029.}
\label{figValidation2}
\end{figure}

The results of the comparison with Horizons are shown in Fig. \ref{figValidation2}. The
discrepancy after 100 days, right before the close encounter with Earth, is of 0.3 m. After that
the discrepancy increases during the close encounter and ends at 3.9 km. The same test in ASSIST
shows a final discrepancy of 500 m against the minor body integrator used at JPL, not
with respect to Horizons. In that test the discrepancy is stable during the first 100 days, around the cm
level. In the present work it is evident that the discrepancy increases linearly in logarithm scale,
suggesting that there is a little discrepancy in the force model with respect to Horizons, but it was
not possible to identify this discrepancy by indirect numerical tests. However, this little
difference, not present in the long-term test, affects only a few meters per year of integration, or few
km after 30 years of integration. As a comparison, the accumulated effect of using the GR method of \citet{dam85},
instead of the PPN formulation restricted to the Sun, is about two orders of magnitude higher. The
conclusion is that this discrepancy does not affect significantly the results of the numerical
integration in SSOXmatch. Tests with asteroids not affected by non-gravitational forces showed the
same situation, suggesting that if there is a difference in the dynamical model, it should be related to a common
force component, and this difference was not found in the long-term test with DE432 ephemerides.

As explained in Sect. \ref{sectionIntegrationSchema}, implicit RK are more suitable for this problem.
However, the comparison with ASSIST, which uses an implicit schema, can be considered quite
satisfactory. Sect. \ref {sectionWebComparison} presents the results of the same test with the Miriade
service, showing that the dynamical model is much more critical than the integrator.

\subsection{Collision of comet Shoemaker-Levy 9 with Jupiter}
\label{sectionPhysModelSL9}

The collision of the fragments of comet Shoemaker-Levy 9 with Jupiter started on July 16, 1994. The A fragment
collided first, officially at 20:13 UTC, although there is an uncertainty of few minutes for all fragments and
the exact time depends on the source. With SSOXmatch the collision is detected 44 minutes before that time,
at 19:29 UTC, with the comet at 69119 km from the position of the barycenter of Jupiter. This distance is
slightly lower than the radius of Jupiter corrected for oblateness (69360 km), considering the fragment entered
from a planetocentric latitude of -45$^{\circ}$. The discrepancy between the barycenter and the physical center is around
150 km, equivalent to a few seconds only in the trajectory of the comet, and this is also the approximate error in
the SSOXmatch estimate, due to the effect of the selected RK, and the time step used in the integration.
With Horizons, a manual iteration shows
the time at which the A fragment had the same distance with respect to the barycenter of Jupiter is around one second
before. Respect the physical center of Jupiter, the correct time is around 2.5 seconds later,
but this correction cannot be computed in SSOXmatch. The ephemerides and cross-matches with HST observations for
other dates show a consistency in the level of 10$^{-5}$ degrees, inline with other results presented in Sect.
\ref{sectionWebComparison}. Table \ref{tableColl} shows the collision times found for all the fragments.

\begin{table}[h]
\footnotesize
\centering
\begin{tabular}{ c c c }
\hline
\textbf{Shoemaker-Levy 9} & \textbf{Collision Julian Day} & \textbf{Distance to Jupiter} \\
\textbf{fragment} & \textbf{(TT)} & \textbf{(barycenter, km)} \\
\hline
F2-A  & 2449550.3123210  & 69118.5 \\
F2-B  & 2449550.5896960  & 69156.0 \\
F2-C  & 2449550.7707451  & 69112.2 \\
F2-D  & 2449550.9663287  & 69272.9 \\
F2-E  & 2449551.1044914  & 69228.4 \\
F2-F  & 2449551.4963932  & 69066.9 \\
F2-G  & 2449551.7864966  & 69097.7 \\
F2-H  & 2449552.2853731  & 69084.4 \\
F2-K  & 2449552.9048999  & 69176.1 \\
F2-L  & 2449553.3996907  & 69184.1 \\
F2-N  & 2449553.9083504  & 69048.9 \\
F2-P2 & 2449554.1110326  & 69063.5 \\
F2-P1 & 2449554.1606055  & 69065.3 \\
F2-Q2 & 2449554.2937926  & 69119.0 \\
F2-Q1 & 2449554.3142466  & 69182.5 \\
F2-R  & 2449554.7038758  & 69238.1 \\
F2-S  & 2449555.1076111  & 69098.6 \\
F2-T  & 2449555.2281448  & 68962.5 \\
F2-U  & 2449555.3879328  & 68952.0 \\
F2-V  & 2449555.6541646  & 69003.6 \\
F2-W  & 2449555.8088213  & 69139.7 \\
\hline
\end{tabular}
\caption{Instants computed for the collision of the fragments of comet Shoemaker-Levy 9 with Jupiter.}
\label{tableColl}
\end{table}

As mentioned previously, to reproduce the results of Horizons these computations should be applied with the
oblateness of Jupiter disabled. With the oblateness enabled the result is around 15 seconds later for the
A fragment.

\subsection{Comparison of orbital elements between the JPL and Lowell databases}
\label{sectionPhysModelJplLowellElem}

\begin{figure}[h]
\centering
\includegraphics[width=1.0\textwidth]{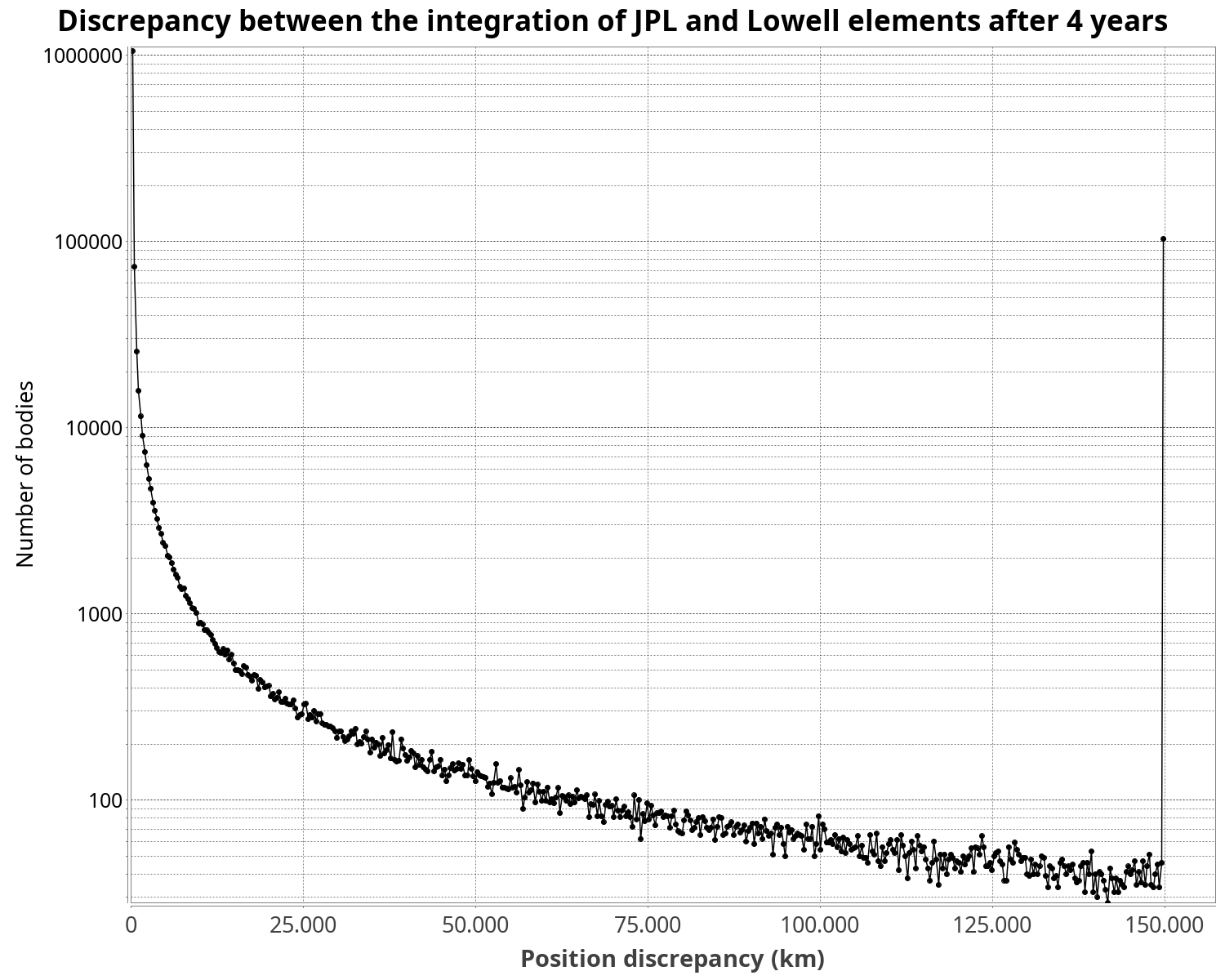}
\caption{Number of bodies showing a given deviation level between the positions returned using JPL and Lowell
elements. The last point shows around 100 000 bodies deviating more than 150 000 km.}
\label{figDeviationElem}
\end{figure}

The comparison of the elements in the two databases is a necessary step to check if there could be any particular
advantage on using one of the two databases. For this purpose a comparison was made on the most recent set of elements
for both databases, performing an integration back in time on more than 1.4 million bodies. The integration was tested
between 1 and 30 years in the past, without any significant difference in the results with respect to the integration period
used, since most of the differences are present in the elements themselves. For reference, the results presented in
this section were obtained with an integration of four years back in
time. A cut limit was set on 15 000 km, or 10$^{-4}$ au, to consider the difference to be significant. This happens
in about 150 000 bodies, 10\% of the total number. This set of bodies, represented in Fig. \ref{figDeviationElem},
can be grouped in different categories:

\begin{itemize}
 \item Bodies discovered and last observed long ago: this is the case of 1979 XB, 1942 RH, 1939 RR, 1935 UZ,
 1988 RH9, and 1927 LA. There are few observations of these asteroids, and the orbits have considerable uncertainties.
 They can be considered as lost bodies, with semimajor axes between 2.2 and 3.3 au. They have eccentricies below
 0.3, except for 1979 XB, with an eccentricity of 0.7. Another body that could fit this category is the
 asteroid Damocles, with a semimajor axis around 12 au, and an eccentricity close to 0.9, that was last observed
 in 1992. This body is highly perturbed by giant planets, making it a nice test for numerical integrators. The
 first four bodies showed huge discrepancies, above 100 million km, the next two bodies were around 0.65 and
 0.37 million km, and Damocles about 0.22 million km.
 \item Around 3400 bodies located in the Kuiper belt, with semimajor axes generally above 40 au, showing
 discrepancies as large as some thousand million km. As an example, the first body of this kind discovered, Albion,
 showed a discrepancy of 100 000 km. These bodies are very faint, so few observations are generally available for
 them, and the orbit uncertainties can be significant.
 \item Around 12 500 bodies with a semimajor axis beyond the Earth, and the perihelion closer to the Sun. 2010 DG77
 is the most representative body, showing a position discrepancy of almost 3$\times$10$^{9}$ km. The JPL database
 mentions a semimajor axis of 2.7 au and an eccentricity of 0.65, while the Lowell database mentions 20.2 au and 0.95.
 For this body the MPCORB database from the MPC provides elements very close to those of the Lowell database.
 \item Bodies with eccentricity below 0.3, and semimajor axes between 2.5 and 4 au. Four bodies belong to numbered
 asteroids, showing a discrepancy around 30 000 km (2008 GO98, 2008 AO68, and 2006 UL41) or more than 100 000 km
 (2015 RZ252). These bodies have been
 observed repeatedly in the last decades, so their orbit uncertainties should be minimal. The vast majority are unnumbered asteroids,
 around 140 000.
\end{itemize}

To perform the test a set of observations were retrieved from the Minor Planet Center for some representative bodies,
with the purpose of reproducing the original astrometry with both sets of elements. The list of observations is shown
in Table \ref{tableObs}.

\begin{table}[h]
\footnotesize
\centering
\begin{tabular}{ c c c c c }
\hline
\textbf{Observation} & \textbf{Body} & \textbf{Observation date} & \textbf{Observatory} & \textbf{J2000} \\
\textbf{number} & & \textbf{(UTC)} & \textbf{code} & \textbf{Position} \\
\hline
1  & 1979 XB    & 1979/12/11.51059  & 413 & 04 26 03.91, -29 29 25.3 \\
2  & 1942 RH    & 1942/09/05.86075  & 062 & 22 55 30.25, -00 30 10.2 \\
3  & 1939 RR    & 1939/09/15.85214  & 012 & 20 39 46.77, -17 45 36.4 \\
4  & 1935 UZ    & 1935/10/19.94893  & 012 & 01 38 12.46, -01 41 33.8 \\
5  & 1988 RH9   & 1988/09/01.01840  & 809 & 21 03 36.26, -11 07 09.0 \\
6  & 1927 LA    & 1927/06/01.96625  & 024 & 16 38 52.58, 01 25 38.4 \\
7  & 2008 GO98  & 2008/04/08.43659  & 691 & 15 42 13.46, -09 41 03.4 \\
8  & 2008 GO98  & 2016/05/08.30454  & 703 & 14 32 17.42, -03 03 14.1 \\
9  & 2008 GO98  & 2024/09/21.215746 & T08 & 16 06 25.493, -07 38 57.59 \\
10 & 2008 AO68  & 2006/10/28.25337  & G96 & 01 21 42.62, 10 05 30.1 \\
11 & 2006 UL41  & 2005/07/07.41683  & 568 & 19 59 35.22, -18 46 28.7 \\
12 & Damocles   & 1991/02/18.52973  & 413 & 09 35 54.96, -72 19 02.1 \\
13 & Albion     & 1992/08/30.46089  & 568 & 00 01 12.79, 00 08 50.7 \\
14 & Albion     & 2008/10/28.356373 & 645 & 01 21 39.39, 09 33 19.2 \\
15 & Albion     & 2022/12/26.196470 & G37 & 02 34 09.995, 16 42 44.91 \\
16 & 2010 DG77  & 2010/02/18.911121 & 097 & 10 44 13.10, -78 40 30.1 \\
\hline
\end{tabular}
\caption{List of observations taken as a test for both the SSOXmatch integration and the JPL/Lowell databases.}
\label{tableObs}
\end{table}

The numerical integration in SSOXmatch was executed using the most consistent options to reproduce the results of the
integration performed by Horizons: 16 perturbing asteroids, the J2 oblateness term of the Earth only, the PPN
formulation of GR restricted to the Sun, the BS54 RK solver, and the Kahan compensanted
summation algorithm. The vectors of the planets were replaced with the output from the JPL DE440 ephemerides after
each integration step. The integration was performed independently with JPL and Lowell elements, and also checked
against Horizons. A summary of the results and their discrepancies with respect to the observations in Table \ref{tableObs}
are presented in Table \ref{tableObsResults}. According to \citet{mos22}, the options selected should be also
compatible with the Lowell database. In Sect. \ref{sectionWebComparison} a complementary table is presented
with the results of the same integration using the Miriade service.

\begin{table}[h!]
\footnotesize
\centering
\begin{tabular}{ c c c c }
\hline
\textbf{Observation} & \textbf{Horizons discrepancy} & \textbf{SSOXmatch discrepancy} & \textbf{SSOXmatch discrepancy} \\ 
\textbf{number} & \textbf{(total and in Ra, Dec)} & \textbf{(JPL elements)} & \textbf{(Lowell elements)} \\ 
\hline
1 & 1.47 (-1.4, 0.3) & 1.47 (-1.4, 0.3) & 2.61 (-2.6, -0.1) \\
2 & 0.00 (0.0, 0.0) & 0.18 (-0.1, -0.1) & 2.99 (2.8, 0.9) \\
3 & 0.14 (0.1, 0.0) & 0.14 (0.1, 0.0) & 0.30 (-0.3, -0.1) \\
4 & 0.36 (0.3, -0.2) & 0.34 (0.1, -0.3) & 0.36 (-0.3, -0.2) \\
5 & 0.15 (0.1, 0.0) & 0.15 (0.1, 0.0) & 0.00 (0.0, 0.0) \\
6 & 0.00 (0.0, 0.0) & 0.15 (-0.1, 0.0) & 0.18 (-0.1, 0.1) \\
7 & 0.00 (0.0, 0.0) & 0.00 (0.0, 0.0) & 3.11 (3.1, -0.2) \\
8 & 0.32 (-0.3, -0.1) & 0.32 (-0.3, -0.1) & 3.08 (3.0, -0.7) \\
9 & 1.55 (-0.6, -1.4) & 1.55 (-0.6, -1.4) & 1.37 (-0.6, -1.2) \\
10 & 45.97 (42.8, 16.7) & 45.97 (42.8, 16.7) & 2.98 (3.0, -0.4) \\
11 & 0.52 (0.1, -0.5) & 0.52 (0.1, -0.5) & 0.30 (0.0, -0.3) \\
12 & 1.32 (1.0, 0.8) & 1.32 (1.0, 0.8) & 1.02 (0.8, 0.6) \\
13 & 0.96 (-0.7, 0.6) & 0.96 (-0.7, 0.6) & 1.20 (-0.9, 0.8) \\
14 & 0.20 (0.0, 0.2) & 0.20 (0.0, 0.2) & 0.36 (-0.3, 0.2) \\
15 & 0.30 (-0.1, 0.3) & 0.30 (-0.1, 0.3) & 0.30 (0.1, 0.3) \\
16 & 38.22 (6.3, -37.7) & 38.22 (6.3, -37.7) & 27.39 (4.6, -27.0) \\

\hline
\end{tabular}
\caption{Position discrepancies in arcseconds with respect to the list of observations in table \ref{tableObs}, computed
by Horizons and SSOXmatch. The total angular discrepancy is shown first, followed by the
individual ones in right ascension and declination.}
\label{tableObsResults}
\end{table}

As a general result, all the astrometry is well reproduced with both sets of elements, despite of providing very
different vectors in recent years. Even for bodies observed over many years, like 2008 GO98, the position uncertainty
is significant. Another visible result is that the output of SSOXmatch with JPL elements matches
Horizons most of the times, and when not most of the discrepancy is generated in the topocentric correction, not in
the integration. Apparently, there is no particular advantage on using any of the two databases. But an important
result is that around 15\% of the unnumbered asteroids have so poorly known orbits that they can be considered
as lost bodies. As a consequency of this, it is very important in a cross-matching software to correctly estimate
the position uncertainty of the bodies, providing this uncertainty in the output, and also an option to skip the
calculation for those bodies with an excesive position uncertainty. The propagation of positional errors is discussed
in Sect. \ref{sectionCEU}.

\subsection{Comparison with the dynamical model used for SkyBoT and Miriade}
\label{sectionPhysModelCompareSkyBoT}

Specific comparison tests between SSOXmatch and the tools implemented by the LTE, for the results
of computing cross-matches and ephemerides, are presented in Sect. \ref{sectionWebComparison}. Here
it is relevant to compare the results of the pre-integration between the dynamical model
described above, and another equivalent to the one used by \citet{ber06}, in which the Earth and Moon are
considered as a single body (the Earth-Moon barycenter, obviously with no treatment of the
oblateness), Pluto is not considered, there are no perturbing asteroids, and their
non-gravitational parameters are not considered. The accuracy of this simplified model when
propagating the elements years into the past strongly depends on the possible presence of
interactions not taken into account, for instance bodies approaching
the Earth-Moon system, or any of the sixteen most massive perturbing asteroids that are
currently considered to develop the JPL and Lowell databases. 

\begin{figure}[h!]
\centering
\includegraphics[width=1.0\textwidth]{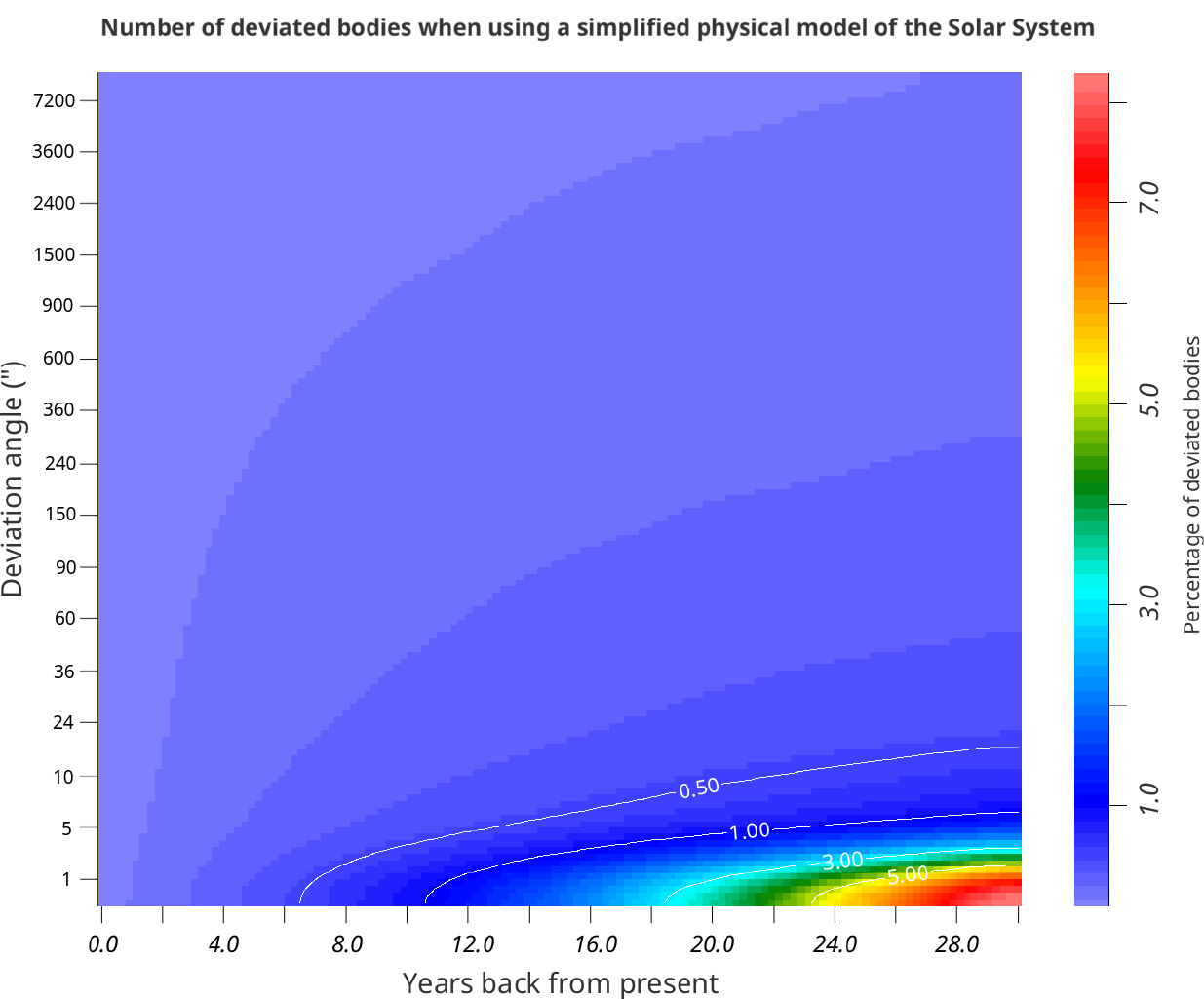}

\caption{Percentage of the total total number of asteroids (around 1.4 million) deviating a given angle
after a given integration period, when comparing positions with a simplified dynamical model of the
Solar System.}
\label{figDeviation}
\end{figure}

The results are presented in Fig. \ref{figDeviation}, that shows the percentage of
asteroids from the Lowell database for which the positions deviates a given angle
after a given integration time. The data was constructed by considering
the minimum distance of the bodies to the Earth during an interval of $\pm$ 1 year around each point in the horizontal
axis (the chart was constructed with a sampling of two years in the horizontal axis, and later interpolated to
artificially improve the resolution), so it represents the number of bodies that will deviate the angle specified in
the vertical axis, at some moment within one year around the instant given in the time axis, specially when they are
close to the Earth. Looking at the numbers, after 30 years
of integration 10\% of the bodies can deviate around the arcsecond or more. The
percentage goes down to 1.3\% for a deviation of five or more arcseconds, and 0.7\% for ten
arcseconds. There are around 1600 bodies potentially deviating 0.5$^{\circ}$ after 30 years of integration,
and around 1000 deviating two or more degrees. After only ten years of integration, the
number of asteroids that could deviate at these high levels is still half of those after 30 years.

As reported in the previous section, the dynamical model used in SSOXmatch is similar to the one used
for the JPL small body database, and to construct the asteroid database of the Lowell observatory. The use of a similar
model is mandatory to reproduce positions close to the observations used to fit
the orbits, and hence to reproduce correct cross-matches when the bodies were visible. Frequently,
this will require years of almost useless integration in which the object
was not observable, but may deviate if a different model is used. The greatest discrepancies observed
were confirmed with Horizons as arising in the simplified
model of \citet{ber06}, affecting specially a number of critical bodies like the NEOs
2020 CD3 or 2015 TC25, that can show wrong positions in SkyBoT or Miriade, by some degrees
(see Sect. \ref{sectionWebComparison}). It is also important to note that in 2006 there
were 360 000 asteroids discovered, and while the relative numbers mentioned above still hold
for deviations of few arcseconds, there is only one body deviating by one degree or more around
year 2000\footnote{This body is the asteroid YORP, the first body for which non-gravitational parameters were
measured.
}
. This shows the model used by \citet{ber06} was adequate for the time,
but since many NEOs have been detected, and the relative number of bodies strongly affected
has increased from 5$\times$10$^{-6}$ to 10$^{-3}$ in 20 years. It is becoming relevant since many studies
are now devoted to NEOs, and in the future it will likely continue increasing.

\section{Cross-matching process}
\label{sectionXM}

\subsection{Space-based observations}
\label{sectionXmIntro}

The computation of cross-matches in SSOXmatch starts from a list of observations taken from one
or several observatories\footnote{For space-based observatories the list is taken from queries to
some public NASA/ESA web archives, see Sect. \ref{sectionWeb}.}. An observation is described by a
polygon defining the area covered in the sky (in astrometric J2000 coordinates), the observation
start and end times (in Universal Time), and the observation identifier. The observations are
sorted from the most recent to the oldest one,
independently from which observatory took it. This list of observations is previously filtered
to remove possible inadequate input data. For instance, some observations may be spread over
a long time, for instance one month, while in the archive the integration time may show only a
few hours. This can be the case of the famous Hubble deep fields. In such cases when we do not
have a detailed information on each individual integration interval, it is probably better to skip
an observation in which the software may return many thousands of possible bodies in the field,
while the probability to find any of them is extremely low. There are two properties to control
this filtering: the maximum allowed time span of an observation (interval between the observation
start and end times, as they appear in the archive), and the maximum allowed ratio between
the previous value and the duration of the observation (exposure time). The values used by default
are 20 days and 10 for these properties.
The maximum acceptable field of view is another property that can be
used to filter big mosaics in which the integration period for each section is not available. The
default value is 20 degrees.



Once the list is clean the computation of cross-matches is executed in a number of
simultaneous threads, all of them working with the same pre-integration file (each of them
covering 73.05 days). For each observation the complete list of around 1.5 million bodies
between asteroids and comets is evaluated, and the cross-matches for each body detected
and registered in a list to later write it to a database or text file. During the process
each body is integrated back in time in little steps from the reference date of the pre-integrated
file read, and the new vectors resulting shared across all threads, so that the integration
needed to check a particular body in a given observation is always short. Numerical integrations are
slow, and to minimize the number of them there are some conditions described
in the next section to filter as much as possible the bodies, by discarding a cross-match in advance, when
possible.

\begin{figure}[h!]
\centering
\includegraphics[width=1.0\textwidth]{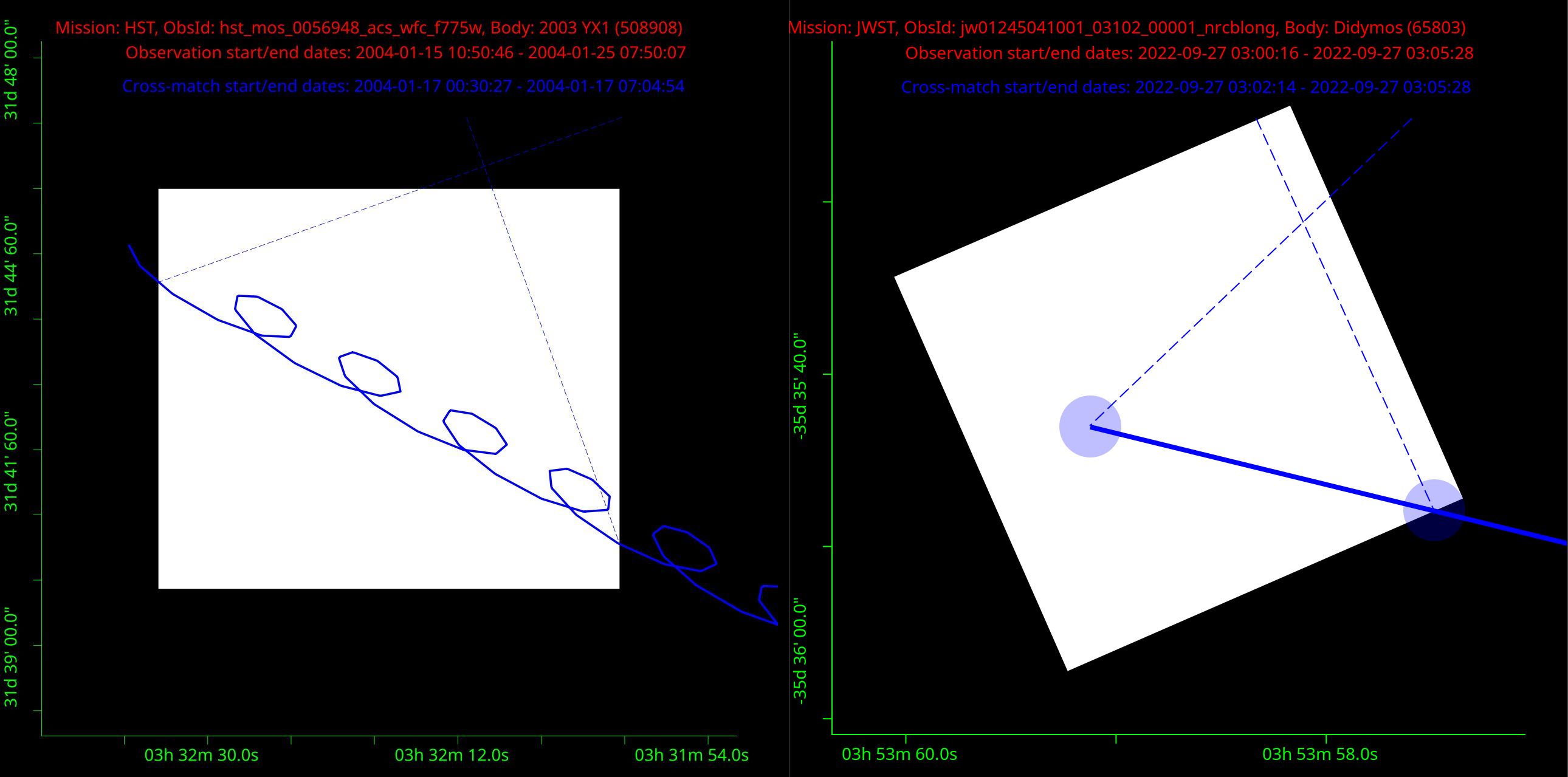}
\caption{Left: Sketch of a cross-match in an HST observation, showing a complex trajectory of the
body due to being very close to HST, and the rotation of the spacecraft around the Earth during
the observation. Right: Sketch of a cross-match of Didymos in a JWST observation, not showing this pattern.}
\label{figXmHST}
\end{figure}

\subsection{General conditions for a particular cross-match}
\label{sectionXmConditions}

One obstacle when computing all possible cross-matches is that any observation is spread over
a time window in which a minor body may move inside and outside the field. In
SSOXmatch the cross-matches are by default computed in a maximum interval of one day,
although the value is configurable. In case an observation is spread over three days,
the cross-matches are computed three times, for each window of one day.
This strategy sets a limit in the maximum movement of the bodies, taken into
account internally during the process.

To accelerate the calculations it is useful to apply all possible speed improvements from
the mathematical and geometrical point of view of the orbits, since in this software
absolute control is possible on the
internal process. For instance, most observations may point to relatively high ecliptic
latitudes, and it is cheap to compute the maximum ecliptic latitude of an asteroid
that is at least one au farther from the Sun than the Earth, corrected for having
the telescope around the orbit of the Earth. When the maximum ecliptic latitude, which
is typically a few degrees like the orbit inclination, is lower than
the minimum ecliptic latitude of the observation (in absolute value), the asteroid will
never appear in the field of view of that observation. Another
strategy is to pre-calculate the solution of the Kepler equation for a grid or values
of eccentricity and mean anomaly, to get a very fast approximate distance of the minor
body, in its orbit and for the time when the observation was taken, to the pointing
direction of the observation. The grid is constructed so that this error will never reach
3 arcminutes. The criteria is to discard the minor body if the distance is greater
than the radius of the
observation plus 10$^{\circ}$ (to account easily for any possible gravitational interactions
that may affect the orbit since the reference time of the elements). In the strange
case the object is closer than 0.1 au to the telescope, or the minor body has
non-gravitational parameters, this criteria is not applied. The reasoning is that these
gravity perturbations will never reach 10$^{\circ}$ in an interval of $\pm$ 73.05/2 days,
even for objects at 0.1 au from the telescope.

If none of the previous conditions obey, the orbit is solved accurately. Another
condition to quickly discard a body is applied later, and is related to the accurate
distance of the body to the field of view of the observation, and the speed of the
object. For bodies located farther than 0.5 au from the telescope, and an observation
radius less than
3$^{\circ}$, the object is discarded if the distance to the center of the field
exceeds 18$^{\circ}$, since it is impossible for any body located far to move such angle in the
sky in less than one day.

In practice, for most bodies one of these conditions will apply, so during the computation of
cross-matches in a given observation, most of the asteroids and comets are discarded
without the need of any expensive operation. Indeed, in modern computers the cross-matches
for a given observation can be computed usually in a fraction of a second, see
Sect. \ref{sectionWebXmAllParam} for a description of one of the web services implemented,
aimed to provide such functionality. Even the cross-matches of a particular body in any
of the thousands of observations of a given survey can be computed in few seconds, see
Sect. \ref{sectionWebXmSingleBody} for a description of the service implementing this
functionality.

When a quick discard is not possible, the cross-match is confirmed only in case the body is found
inside the field of view, whatever its shape. For that, the sky coordinates of the field
are transformed to pixel coordinates using an stereographic sky projection, like if
the intention is to draw everything on the screen. Then, a Java object is constructed
to represent that shape, and native Java functions are used to calculate if the body
is inside the shape or not, applying all necessary time subintervals to find if there
is a cut along the trajectory for a particular instant (performing numerical integration in these
subintervals when the object is close to the telescope), or if the position plus the
uncertainty lies partially within the field. This direct approach is different from the
approach considered in other pipelines based on a software designed to compute ephemerides,
not specifically to search for cross-matches. In \citet{rac22} the ephemerides library
Eproc is used, which is also the library behind the SkyBoT and Miriade services. This
cross-matching pipeline calls Eproc in a two-step process, using
first the HEALPix sky tessellation technique to calculate
globally the positions of all bodies (despite after the numerical
integration most would never appear on the field), and then, in a second step, the
positions of the cross-matching candidates
are re-computed with a shorter time step to confirm or discard the cross-match, using a set of
geometrical algorithms to interpolate in-between. There are three types of cross-matches in
\citet{rac22}: type 2 corresponds to objects found inside the field, type 1 to bodies for
which the position plus the positional error lies partially within the field, and type 3 corresponds
to bodies for which this interpolation between the closest two positions computed cut the
observation footprint.
It is noticeable that for some surveys like HST it is not possible to ensure the body
will follow a straight line during an interval of time, allowing such interpolation,
since the HST itself moves in a curve around the Earth (see the sketch of the cross-match
example in Fig. \ref{figXmHST}).

In SSOXmatch there is a cross-match type field following the same criteria,
but the type 3 does not exist in this software since all cross-matches are confirmed
with a cut of the field of view. The approach of using pixel coordinates prevents issues
similar to those possible in the interpolation described above, when the minor body
is close to the poles.

\subsection{Propagation of positional errors}
\label{sectionCEU}

In the Lowell database there are some fields aimed to help to compute the position uncertainty of the
asteroids, the so called CEU parameters. They are expressed
in units of arcseconds for the current 1-$\sigma$ position uncertainty (referred to a date close to
that of the elements, which is evolving continuously), and a rate of change in arcseconds per day,
to be added to get the final in-orbit uncertainty for a given date. But the real uncertainty for the purpose
of the present work depends on if the body is very close to the telescope, or very far, so the in-orbit
uncertainty should be corrected for the position of the telescope. This method is very simple to apply,
and works well \citep{des13}, but may fail once the body has suffered a close encounter with the Earth
or any other massive planet. In addition, it is strictly an in-orbit uncertainty, which does not account
for the uncertainty in the orbital elements pointed out in Sect. \ref{sectionPhysModelJplLowellElem}, for
instance those mentioned for 2010 DG77.

On the other hand, the JPL database offers a wide variety of different fields, including the 1-$\sigma$
uncertainties in each of the derived orbital elements. During the development the easiest approach was to use a
common, simple method to derive the position uncertainty, using the same model as in the Lowell database.
For this purpose the parameters condition code, fitting error, and the middle point of the
arc of time the body was observed was used to estimate a position uncertainty. The condition code in the
JPL database is the same as the \href{http://www.minorplanetcenter.org/iau/info/UValue.html}{U parameter defined by the Minor Planet Center}
for the MPCORB database. From this parameter the rate
of change of the CEU can be derived within a factor two. For the date of CEU the criteria implemented for
JPL elements is to use the midpoint of the arc of time the body was observed, using the dates provided for
the first and the last observation used in the fit. For the CEU uncertainty in this date, the value assumed is the 1-$\sigma$
uncertainty of the orbit fitting, scaled with the distance to Earth on the CEU date and the length of the orbit
path, to estimate the in-orbit position uncertainty. This is based on the fact that an Earth-based
observatory is always the origin of, at least, most of the astrometric observations.

\begin{figure}[h!]
\centering
\includegraphics[width=0.52\textwidth]{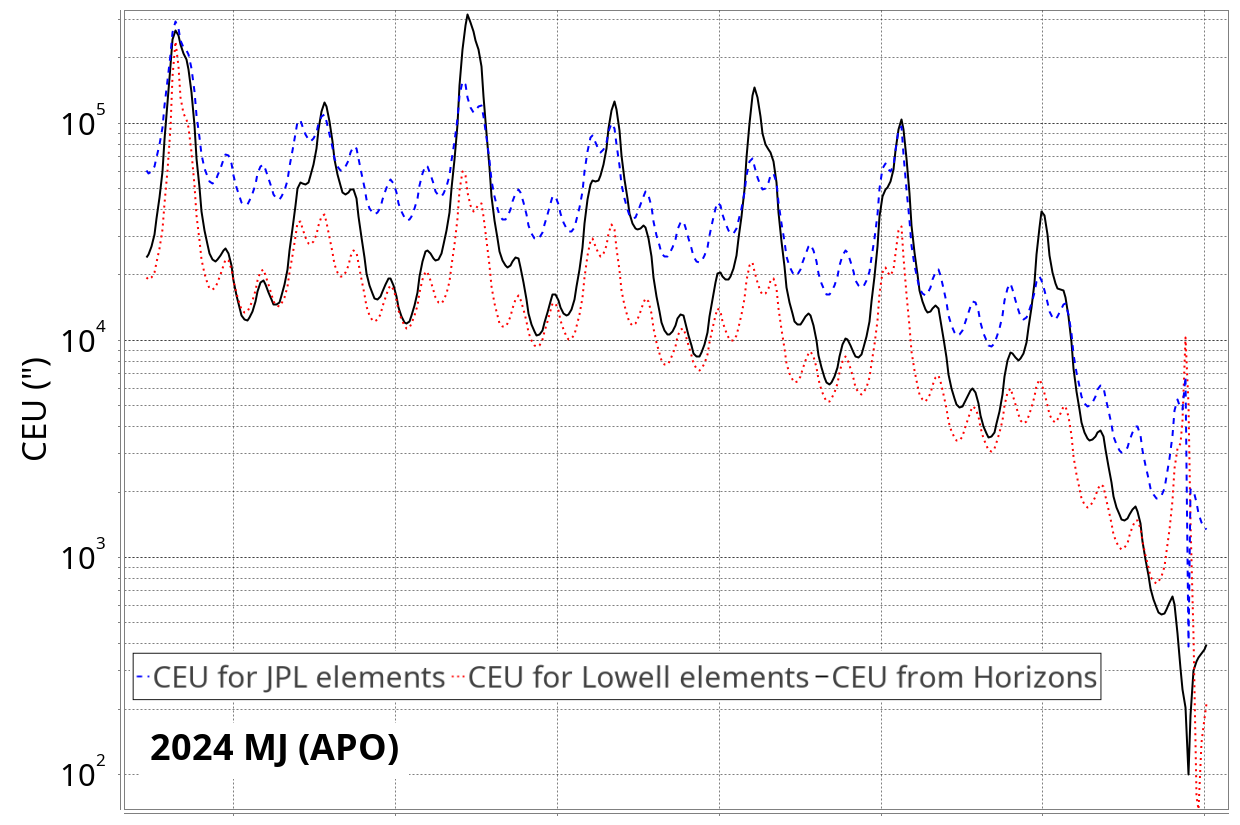}
\includegraphics[width=0.47\textwidth]{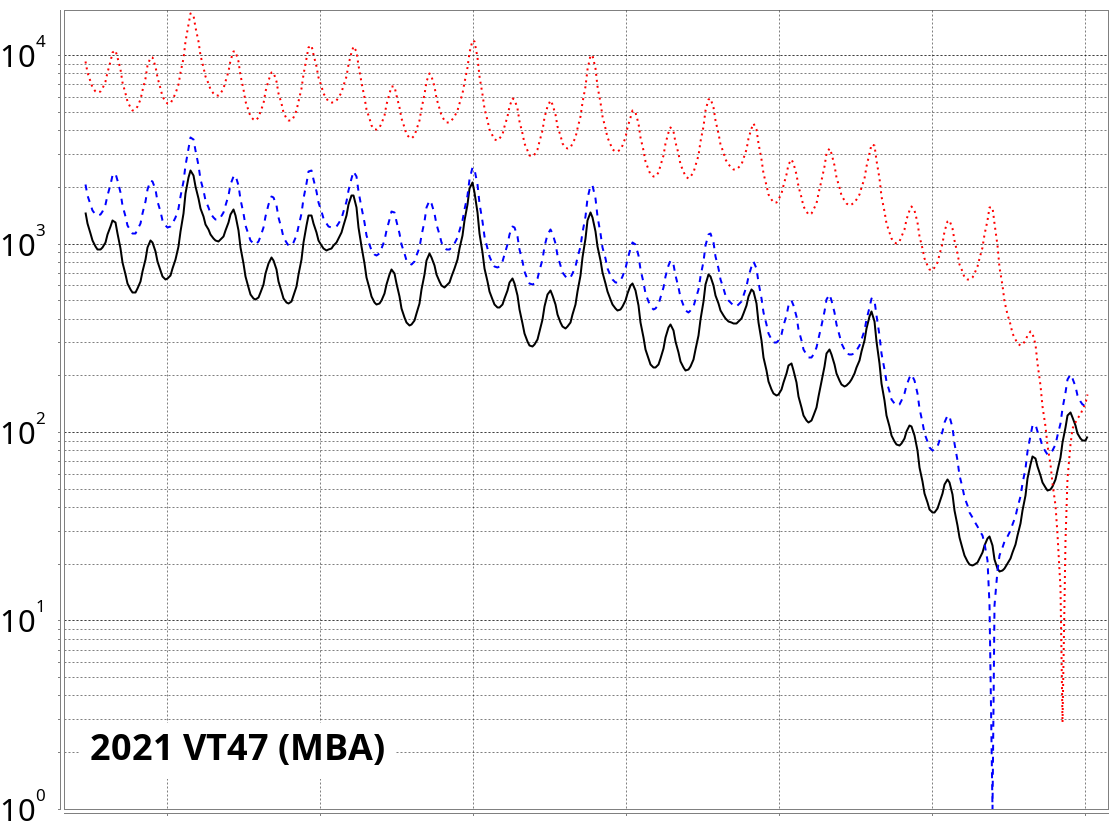}
\includegraphics[width=0.52\textwidth]{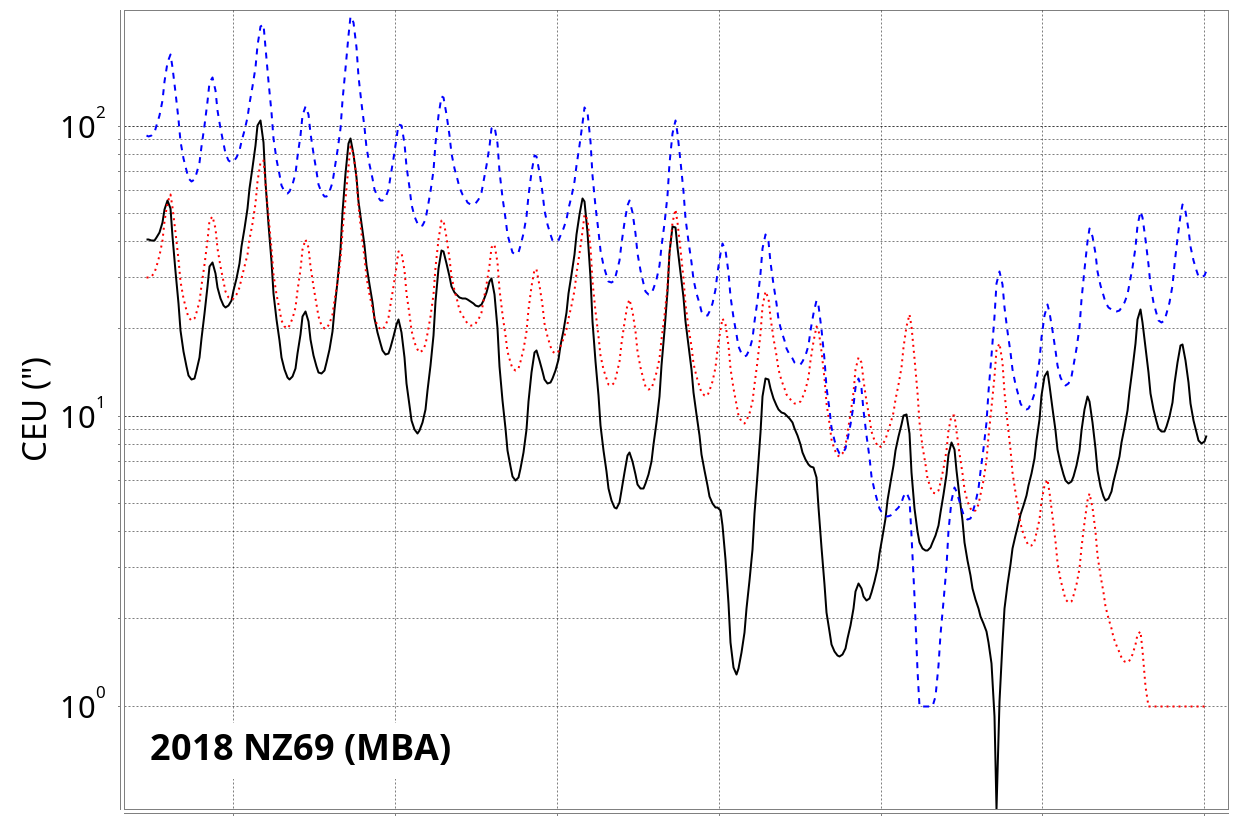}
\includegraphics[width=0.47\textwidth]{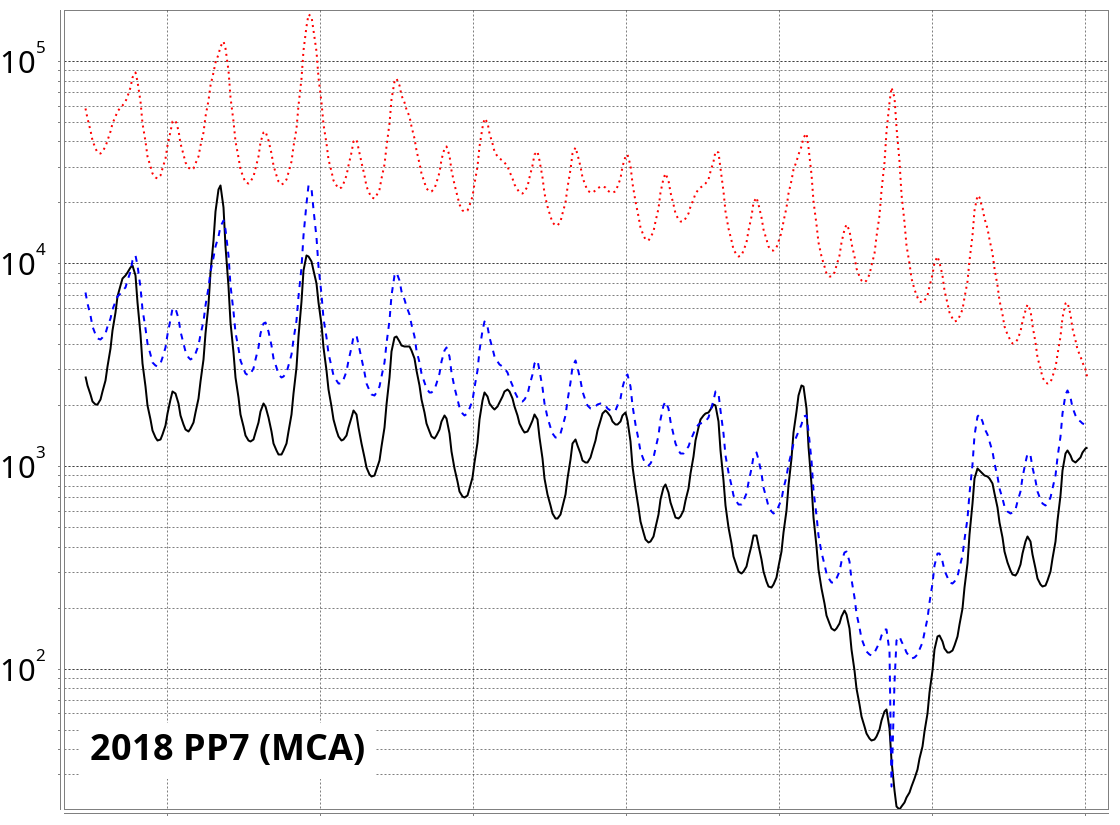}
\includegraphics[width=0.52\textwidth]{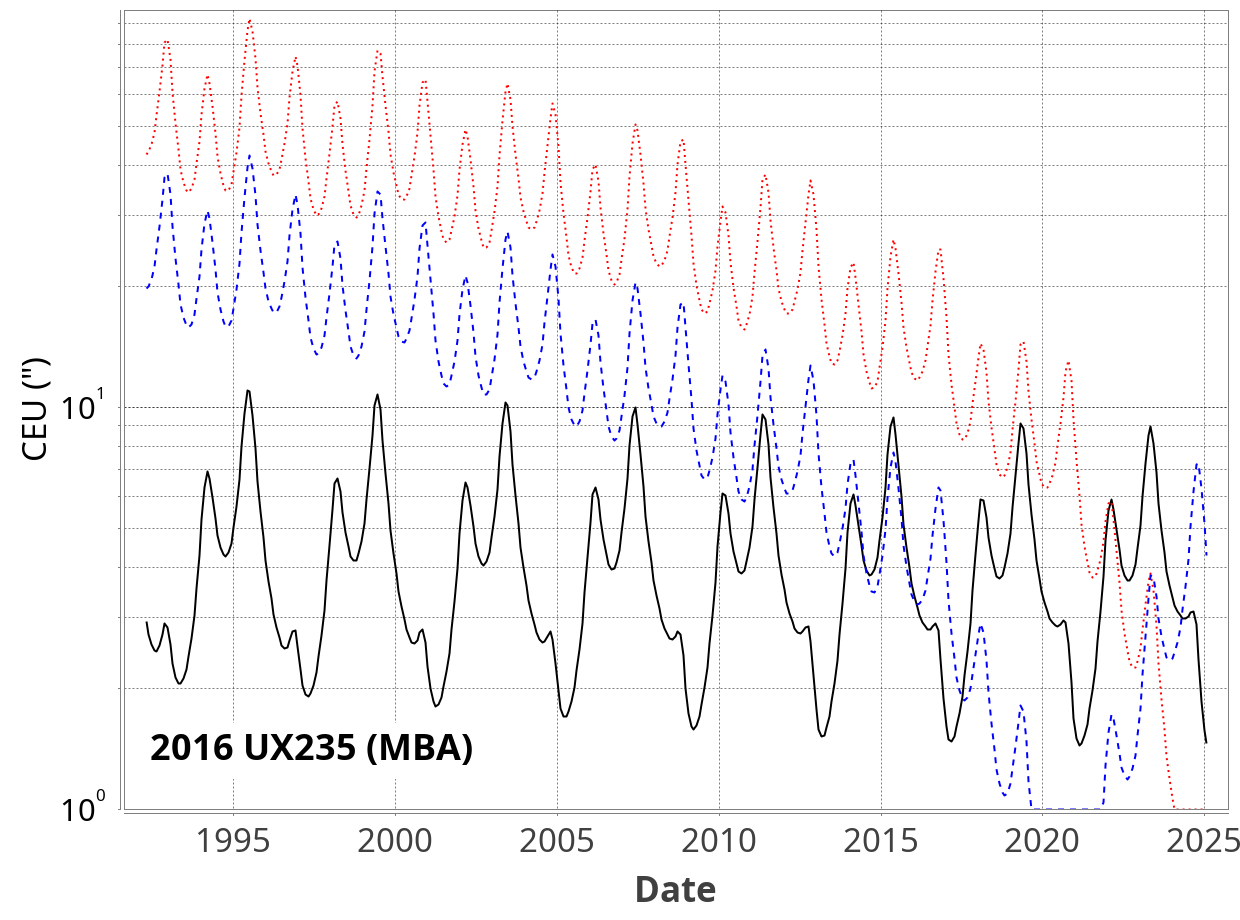}
\includegraphics[width=0.47\textwidth]{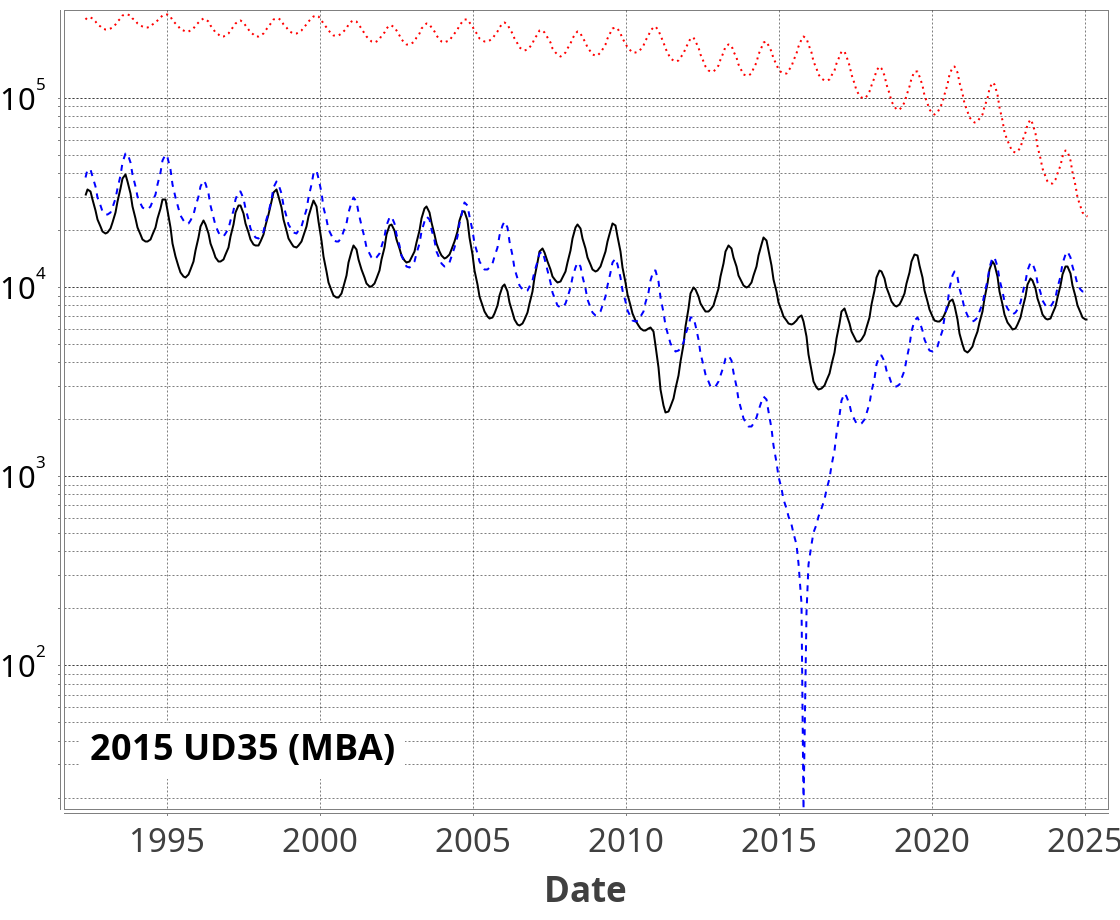}
\caption{Comparison of the positional uncertainty from Earth of some bodies, using the criteria
based on CEU parameters in the Lowell database (red dotted line), the equivalent method developed for JPL elements (blue dashed line),
and the uncertainties reported by Horizons (black line). Asteroid names and their dynamical classes appear in th bottom-left corner.}
\label{figCEU}
\end{figure}

Fig. \ref{figCEU} shows a comparison between the position uncertainty using the parameters in the Lowell
database (red line), the equivalent method developed for JPL elements (blue line), and the uncertainty
reported by Horizons (black line), for some bodies. When the body gets
closer to the telescope, in this case assumed to be at the center of the Earth, the angular uncertainty
shows a peak, except for the dates in which the body was observed, when this value is lower. This means
the appearance of the chart for a given body depends on the (space) telescope observing it.
In general, the agreement is within a factor 5 between all methods, and the method adapted for JPL elements
matches Horizons better than the CEU values in the Lowell database. In some cases, the minimum uncertainty
does not happen around the same dates in the curves, showing that the CEU date assumed may not be the most
adequate value, but it is the only one that can be derived with the information provided in the
small body database.

For comets a fixed value of 10" from the observer is used by default, to cover the possible extension
of the tail, but with no increment with time or in-orbit calculations. Optionally it is possible to add a
percentage of the trajectory correction due to the non-gravitational parameters, to
account for a possible uncertainty in the provided parameters, or a possible change
of them with time. The default value used in SSOXmatch is 1\% of the effect of these
parameters, that may
eventually become higher than 10". In Horizons it is possible to find different
sets of elements for different orbit numbers of some comets, to account for this or
other kind of orbit changes, but this is beyond the purpose of SSOXmatch.

As conclusion, the position uncertainties computed seems robust enough, although they are based on in-orbit
calculations, assuming the orbit shape is well-defined, which is an incorrect assumption for a considerable
number of asteroids, as mentioned in a previous section. If the computed position uncertainty of the body
for the date is higher than a value selected in the properties, if any, the object is discarded and no
cross-matches are computed for it. The default value in the properties is
0.01$^{\circ}$ and 10 times the radius of the observation, and both conditions should
obey. If an observation only covers a field of 1" and the error is more than 10",
it is very unlikely to find the object in the image, but in this case probably that
body was the purpose of the observation, so the absolute condition 0.01$^{\circ}$ would
allow the cross-match to be saved, unless the uncertainty is also above this
value. So this value should not be very low, specially for comets and for asteroids
that were observed over a limited arc of time.

\section{Cross-matching results for the space surveys}
\label{sectionXmResults}

There are papers that tackle the identification of asteroids from the trails detected in the
observations, see for instance \cite{kru21} for HST observations. The purpose of this section is different: to
present a complete catalog of all known Solar System bodies that may be identified in the complete set of
observations from HST, and others surveys as well: JWST, XMM, Spitzer, and Herschel. The list will be filtered
considering the limiting magnitude of the telescopes and the position uncertainty of the ephemerides of the
asteroids.

\begin{table}[h!]
\footnotesize
\centering
\begin{tabular}{ c c c c c c }
\hline
\textbf{Survey} & \textbf{Number of} & \textbf{Obs. with} & \textbf{Total} & \textbf{Individual} & \textbf{Limiting} \\
\textbf{} & \textbf{observations} & \textbf{asteroids} & \textbf{XMs} & \textbf{bodies} & \textbf{magnitude} \\
\hline
HST  & 4370325 & 135447 & 200579 & 17763 & 26 \\
XMM  & 15002 & 5743 & 36384 & 29190 & 22 \\
JWST  & 959783 & 11518 & 14043 & 1310 & 29 \\
Spitzer  & 43789 & 2736 & 54266 & 11674 & 22 \\
Herschel  & 45422 & 9302 & 620136 & 216205 & 26 \\
\hline
\end{tabular}
\caption{Basic statistics of the asteroids that may show cross-matches in the different surveys, assuming
the limiting magnitude shown in the last column, and a cross-match type 2 (object crossing the field). The
limiting magnitudes assumed for the infrared telescopes Spitzer and Herschel are just informative, since they
heavily depend on the wavelength and the beam dilution factor.}
\label{tableSurveys}
\end{table}

Table \ref{tableSurveys} presents the basic statistics of the five surveys, with the number of observations
considered, the number of them potentially presenting bright enough asteroids, the total number of cross-matches
produced (in most cases more than one body is present in the observation), the number of individual bodies producing
these cross-matches, and the limiting magnitude considered to get the numbers. The ratio between the third and the
second columns gives the percentage of observations that may contain an asteroid trace. For HST this value is 3.1\%,
which is very consistent with the value found by \citet{kru21}, showing that this synthetic study applied to a
catalog of observations provides a similar result than works based on analyzing real observations. There is a
large number of cross-matches and individual bodies for Herschel. The explanation lies on the presence of around
2000 observations listed in the archive with a field of view above the degree, with around 30 of them above five
degrees. There is also a considerable number of mosaics with a size around the degree in the Spitzer archive. The
JWST and HST archives are in the opposite situation. As explained in Sect. \ref{sectionXmIntro}, the observation start/end
times are not applicable to mosaics (this metadata may not reflect the integration interval of the individual sections
of the images), so it could be better to filter them. More accurate data may be present in the header of the fits files.

\begin{figure}[h!]
\centering
\includegraphics[width=1\textwidth]{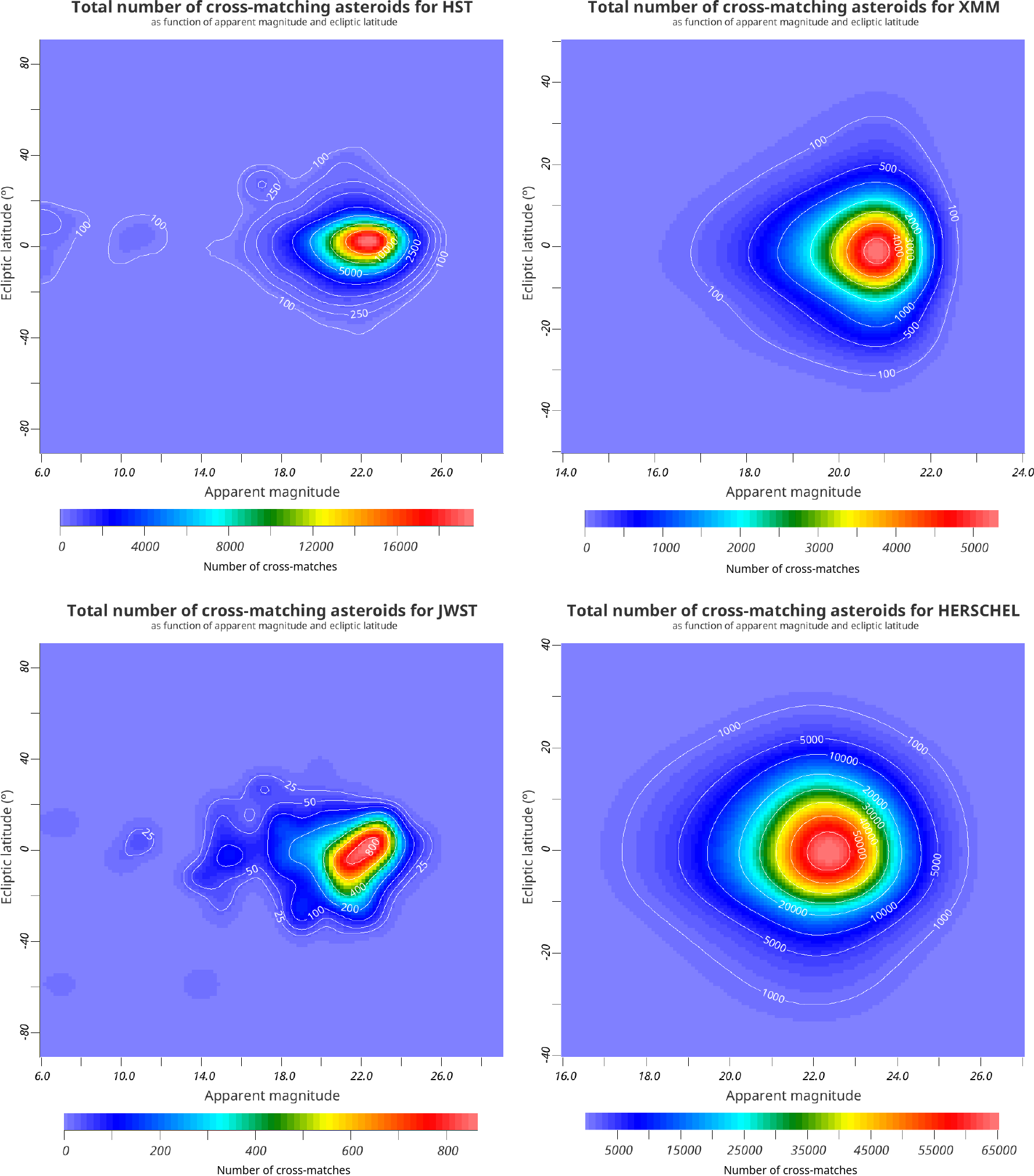}
\caption{Distribution of asteroid cross-matches potentially detectable for different surveys in specific observations taken,
as function of the apparent magnitude and the ecliptic latitude from the telescope.
Contour levels show the number of cross-matches.
}
\label{figSurveys}
\end{figure}

Fig. \ref{figSurveys} presents the number of cross-matches found in the surveys as function of the apparent magnitude
and the ecliptic latitude of the body, as seen from the telescope. Spitzer is not shown since its chart is very
similar to the XMM one. For HST the peak where most cross-matches are located is at magnitude 22.4, and +1$^{\circ}$
of ecliptic latitude. The peak at magnitude 17 and latitude +27$^{\circ}$ belongs to the Kuiper belt body Haumea. In the
region around magnitude 11 there are 17 bright asteroids observed by HST, between (5) Astraea and (3200) Phaeton. The
little peak around magnitude 7 belongs to Ceres and Vesta. For JWST it is notorious that the chart is wider both in
magnitude and latitude, showing the superior capability of JWST to detect faint asteroids. The irregular
shape suggest that some specific asteroid populations can be identified. The peaks around the
-60$^{\circ}$ of latitude belong to Pallas and Didymos. Pallas has an orbital inclination with respect to the ecliptic
above 30$^{\circ}$. The inclination of Didymos is of few degrees, but some JWST observations where taken when this
body was very close to the telescope.

\begin{figure}[h!]
\centering
\includegraphics[width=1\textwidth]{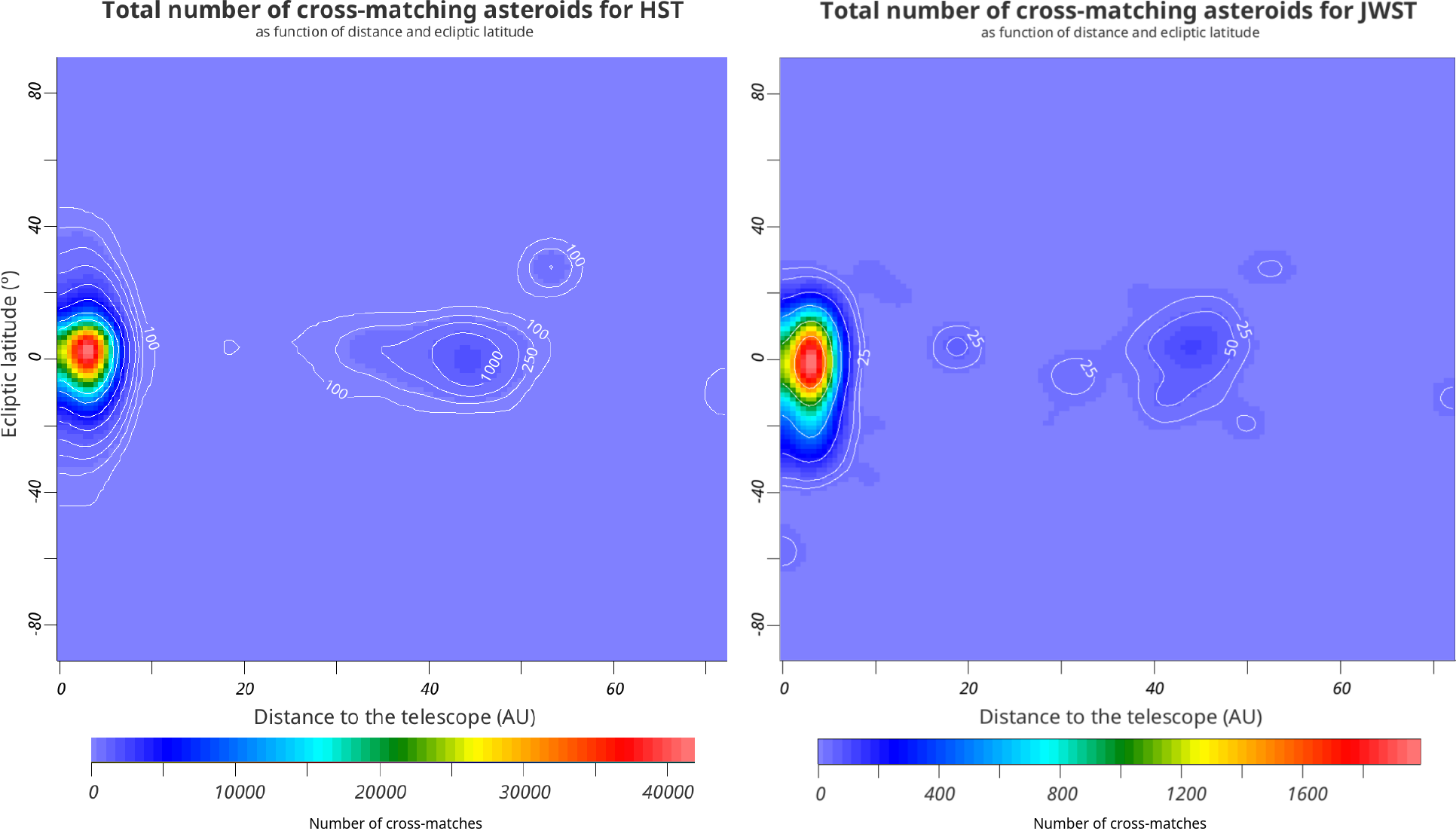}
\caption{Distribution of asteroid cross-matches potentially detectable for the HST (left) and JWST (right) in specific observations already taken, as
function of the distance to the telescope and the ecliptic latitude.
Contour levels show the number of cross-matches.
}
\label{figSurveys2}
\end{figure}

\begin{figure}[h!]
\centering
\includegraphics[width=0.49\textwidth]{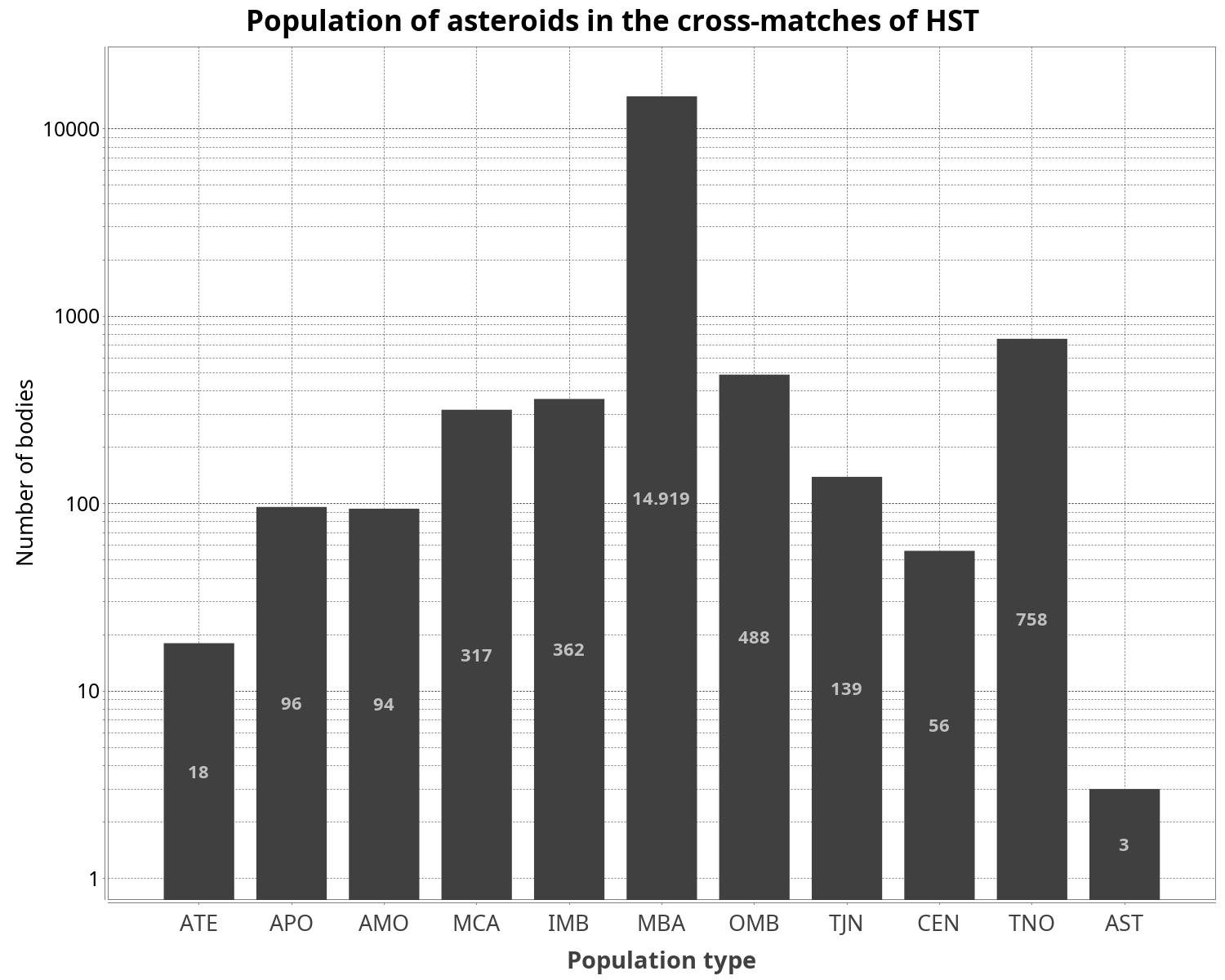}
\includegraphics[width=0.49\textwidth]{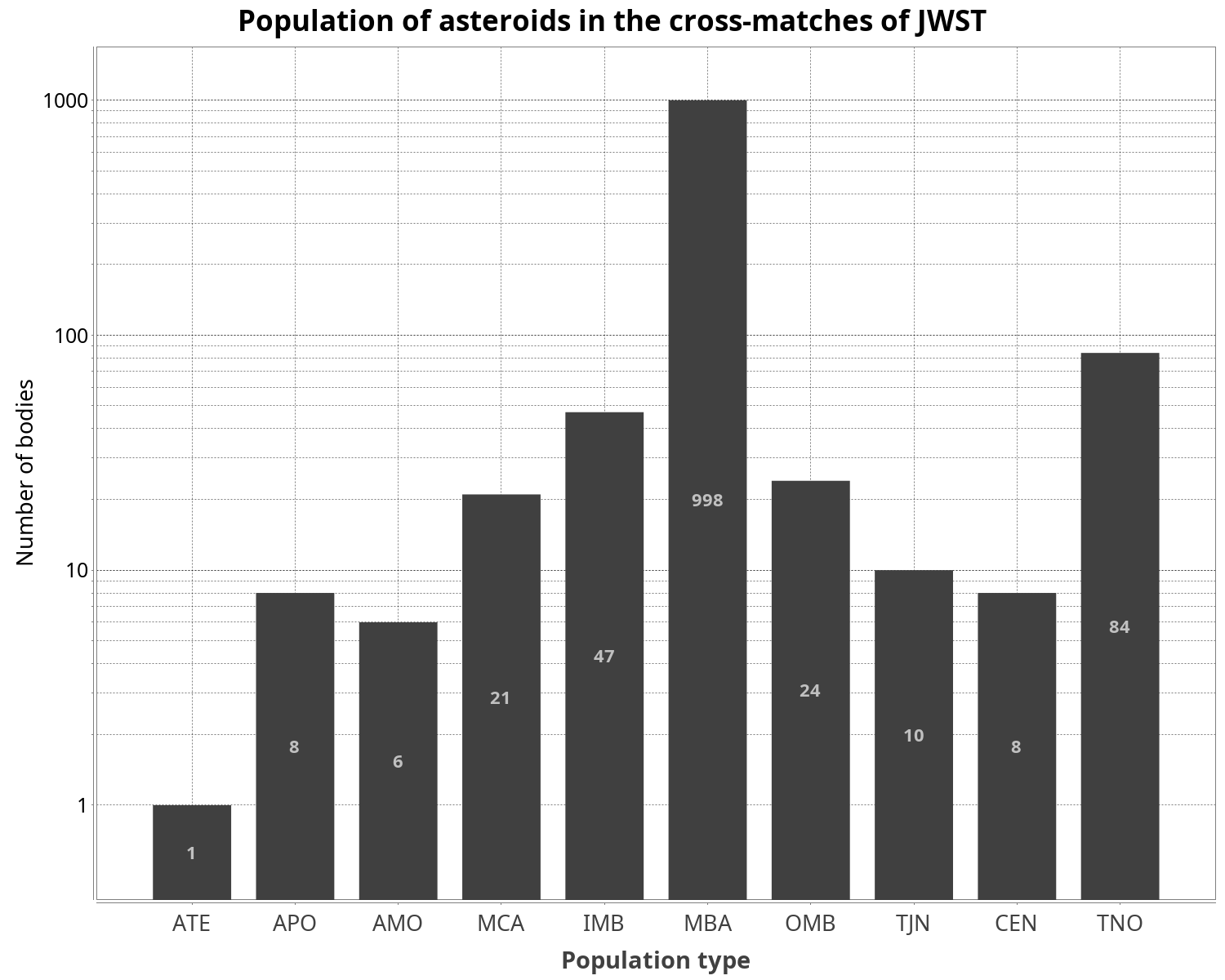}
\caption{Population of asteroids for the potential cross-matches in HST (left) and JWST (right). The orbit classes shown from the
JPL catalog, ordered by semimajor axis, are:
ATE (Aten, belong to NEOs), APO (Apollo, also NEOs), AMO (Amor, also NEOs),
MCA (Mars-crossing Asteroid), IMB (Inner Main-belt Asteroid), MBA (Main Belt Asteroid),
OMB (Outer Main-belt Asteroid), TJN (Jupiter Trojan), CEN (Centaur),
TNO (TransNeptunian Object), AST (Asteroid not belonging to specific family).}
\label{figSurveys3}
\end{figure}

Fig. \ref{figSurveys2} shows the number of potential detections in observations from HST and JWST already taken as function
of the distance to the telescope in the instant of the cross-match, instead of the apparent magnitude. Other surveys are not shown
since the contamination due to mosaics has not been removed. The vertical axis is still the ecliptic latitude.
In the JWST the different populations appear more clearly. There is a little extension in the main spot at 10 au and +24$^{\circ}$,
that belongs to the TNO 2013 LU28, and interesting body with an eccentricity of 0.95, which can be also classified as
a Centaur, with a retrograde orbit. Other Centaurs appearing are Okyrhoe, (32532) Thereus and 2013 XZ8, and
2020 VF1, on
ecliptic latitudes of 0$^{\circ}$, -15$^{\circ}$, and -37$^{\circ}$, respectively. The peak at 19 au belongs to Chiron, and the extension at 31 au to
(47171) Lempo and five additional bodies. After the main group of Kuiper belt objects, we can see the cross-matches
of Makemake at 52 au, and 2005 QU182 at 55 au. Beyond 80 au there are still some observable bodies like Sedna, Gonggong, and Eris,
but they are located outside the limits of the charts. The HST chart is quite similar, but slightly less defined.
The number of cross-matches on it is considerably
higher due to the extension of the mission. Due to this the individual Centaur bodies observed cannot be inferred,
since there are many of them. In Fig. \ref{figSurveys3} the population of asteroids with potential cross-matches with
HST and JWST is shown. They are compatible with the results of \citet{kru21}, although the asteroid classes presented,
taken here from the classification in the JPL database, are not the same.

\section{Web services}
\label{sectionWeb}

\subsection{Services implemented}
\label{sectionServices}

\begin{figure}[h!]
\centering
\includegraphics[width=1.0\textwidth]{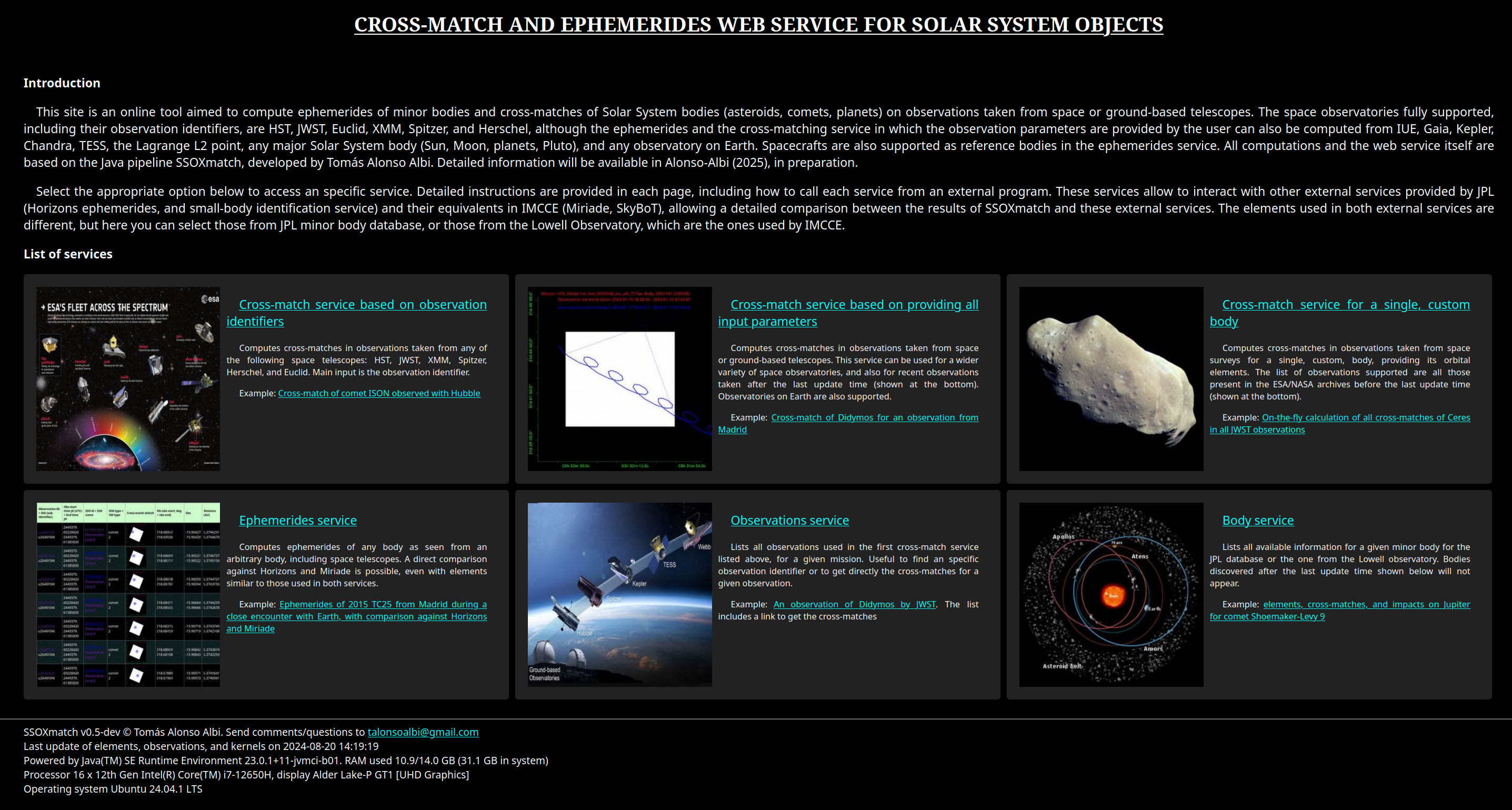}
\caption{Main web page of the services.}
\label{figMainService}
\end{figure}

There are a total of six web services implemented. From them,
the most relevant ones are the cross-matching service based on providing all input parameters
(Sect. \ref{sectionWebXmAllParam}), which is equivalent to SkyBoT and the small body
identification service from JPL, and the ephemerides service (Sect. \ref{sectionWebEphem}),
equivalent to Miriade and Horizons. The main page is shown in Fig. \ref{figMainService}.

The web site is entirely implemented in Java, in the same package SSOXmatch, which is in charge
of presenting the services, performing the requested computations, and returning the output in
a variety of formats, from html tables returned within the same web page, to responses in json or csv
formats, which are more suitable for external scripts or programs. The services are currently not
compliant with the Virtual Observatory, but this evolution will be studied in the future. The
dependencies are updated automatically every week, with incremental updates for new bodies,
observations, and their corresponding cross-matches.


\subsubsection{Cross-matching based on observation identifiers}
\label{sectionWebXmObsId}

\begin{figure}[h!]
\centering
\includegraphics[width=1.0\textwidth]{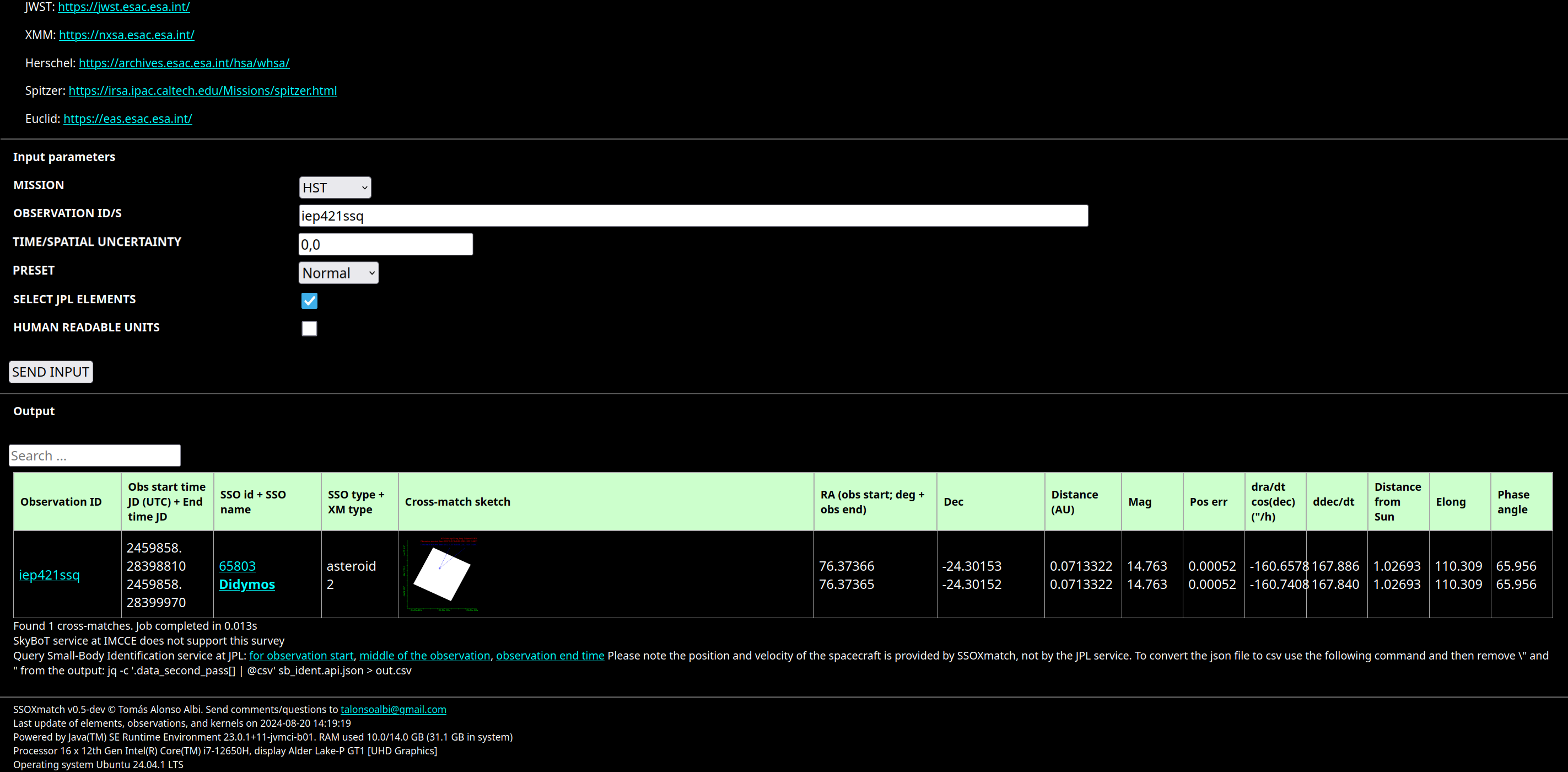}
\caption{Bottom of the page of the service providing cross-matches for a given observation identifier, showing
a cross-match of asteroid Didymos in an observation taken by HST.}
\label{figXmObsId}
\end{figure}

This service provides the list of bodies that may appear in a given observation taken from a supported space
mission. The main input is the identifier/s of the observation/s, the mission the
identifiers belong to, and if JPL elements should be used, or those from the Lowell observatory and
the cometpro database. The limiting magnitude of each survey is set as one of the parameters of the properties
file when the software is launched, so that cross-matches fainter than this limit will not be computed. However,
probably not all cross-matches reported will appear in the observation, since the limiting magnitude may depend
on the instrument, exposure time, or the proper motion of the minor body. As with other
services, detailed instructions are provided at the top of the page, and some of the mentioned options are
common in the other services.

The screenshot of the output from the service is shown in Fig. \ref{figXmObsId}. The ephemerides of the body
are provided for the start and end of the observation, and a sketch of the cross-match is presented,
showing the useful information visually. The sketch can be enlarged by clicking on the image. The
observation identifier and Solar System Object columns are links to other services in the web site,
providing additional information about the observation (see Sect. \ref{sectionWebObs}) and the
body (see Sect. \ref{sectionWebBody}), respectively. Below the table some links are provided to
perform the same computations in SkyBoT and the JPL service, for an easy comparison of the results. The links
are provided for different times from the start to the end of the observation.

The surveys supported include \href{https://hst.esac.esa.int/}{HST}, \href{https://jwst.esac.esa.int/}{JWST},
\href{https://nxsa.esac.esa.int/}{XMM}, \href{https://archives.esac.esa.int/hsa/whsa/}{Herschel},
\href{https://irsa.ipac.caltech.edu/Missions/spitzer.html}{Spitzer}, and \href{https://eas.esac.esa.int/}{Euclid},
for which the cross-matches are limited to a subset of Level 1 observations taken on December, 2023.
The list of observations are publicly available from the mentioned web services using the \href{https://www.ivoa.net/documents/TAP/20190927/index.html}{Table Access
Protocol (TAP) service}, which is a protocol for retrieving data in the \href{https://www.ivoa.net/}{Virtual Observatory initiative} \citep{gen17}.

\subsubsection{Cross-matching based on providing all input parameters}
\label{sectionWebXmAllParam}

This service mimics SkyBoT and the equivalent JPL service. It is similar to the previous one,
but instead of an observation identifier the user supplies all the parameters defining an observation,
like the start/end times, and the footprint in the sky, providing the center position and the radius, or
a polygon defining in detail the shape of the area covered in the sky. The output of this service is
identical to the one based on observation identifiers, shown in Fig. \ref{figXmObsId}. As was mentioned, the
computations of cross-matches over a given time window and specific sky region, which are key
elements in this service, are not supported in SkyBoT or the JPL service. To overcome the limitation
of the observer location in the JPL service, the links provided below the output table to perform the
computations with this service include positions computed by SSOXmatch, using the kernels of the
space telescopes. This workaround is not possible in SkyBoT (see Fig. \ref{figXmObsId}).

The origin from which the observation is taken is specified with an spacecraft/observatory code,
supporting a large list of spacecrafts (HST, JWST, XMM, Spitzer, Herschel, Euclid, IUE, Gaia, Kepler,
Chandra, Tess, or a generic Lagrange L2 point), any major Solar System body like the Sun, Moon, planets,
Pluto, and any IAU code for an observatory on Earth, written as a three-character code.

\subsubsection{Cross-matching for a single, custom body}
\label{sectionWebXmSingleBody}

From a set of orbital elements provided
as input, for instance for a new object discovered, the service can compute the cross-matches of this
specific body in all observations taken from any of the space-based surveys supported. This can be
used to search for a new body in any of the old observations. The numerical integration is performed
entirely on-the-fly, and the observations filtered as explained in previous sections, so that the
computations are completed in few seconds, generally. The entire HST survey, containing millions of
observations spread over decades, can be processed entirely for a single body in about 10 minutes.
A similar, much faster service is provided by the
\href{https://www2.cadc-ccda.hia-iha.nrc-cnrc.gc.ca/en/ssois/}{SSOIS} \citep{gwy12}, for the
observations of a large number of facilities, including the HST space telescope.

\subsubsection{Ephemerides}
\label{sectionWebEphem}

\begin{figure}[h!]
\centering
\includegraphics[width=1.0\textwidth]{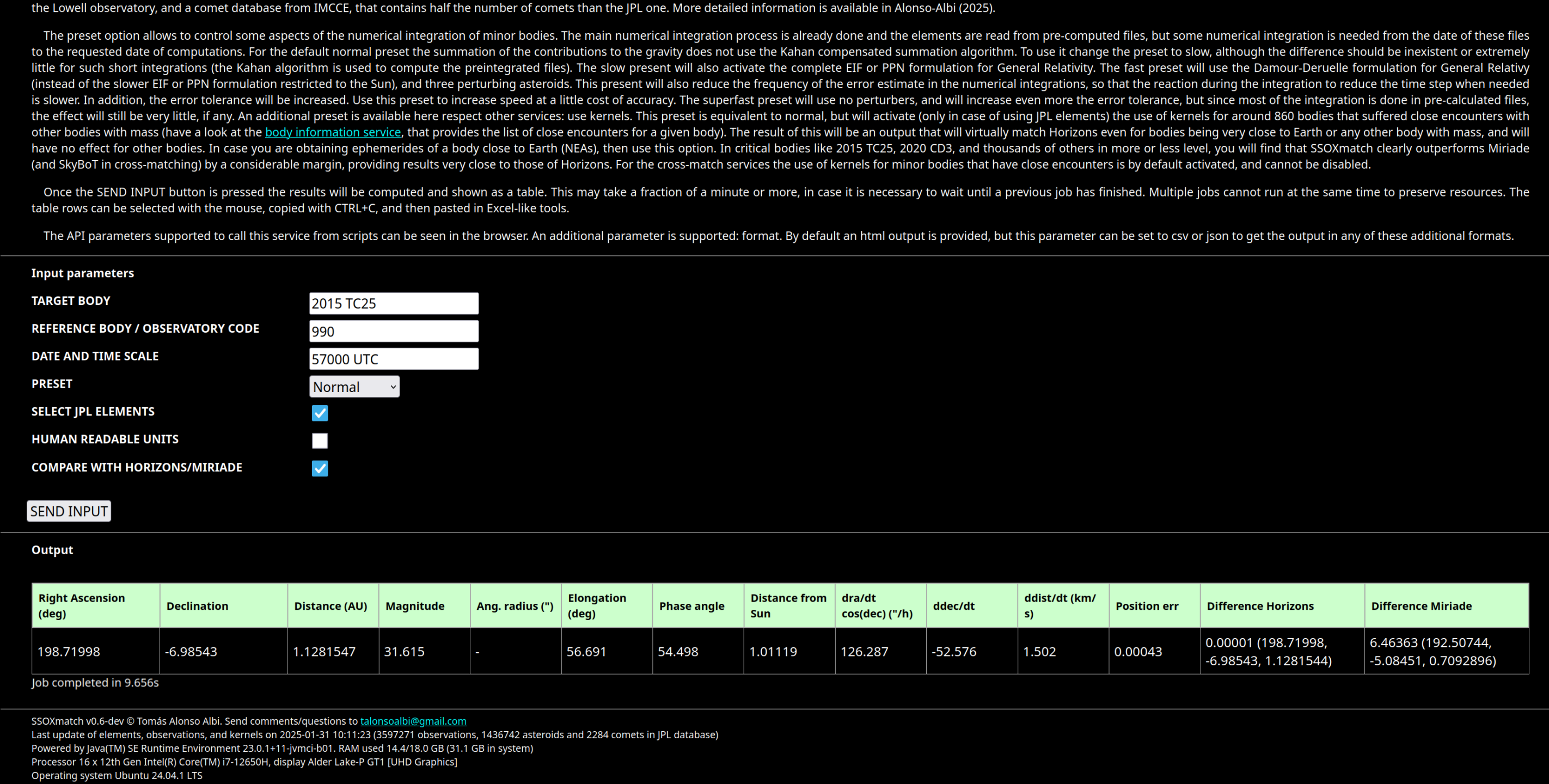}
\caption{Bottom of the page of the service providing ephemerides, showing the ephemerides of 2015 TC25 during a
close encounter with Earth.}
\label{figWebEphem}
\end{figure}

This service is equivalent to Miriade from the LTE, or to Horizons from the JPL websites. The main input parameters
are the target body name, the name of the reference body or observatory code to use as the position for
the observer, the date and time, and the selection between JPL and Lowell elements. The service
provides a link to the list of observatories supported, which is updated automatically, to help the user find
the adequate one. The target
body can be any Solar System body, including satellites and minor bodies, as well as any spacecraft supported
(the complete list was mentioned
in Sect. \ref{sectionWebXmAllParam}). The reference body can be any major Solar System body
(Sun, planets, Pluto, and the Moon), spacecrafts, or the code for any observatory on Earth. This flexibility
in the input is above the level of Miriade, and slightly below that of Horizons. An
screenshot of the service is shown in Fig. \ref{figWebEphem}.

\subsubsection{Observations}
\label{sectionWebObs}

This service simply provides additional information for a given observation identifier belonging
to a space-based observatory. The identifier filter is optional, allowing to list all observations for a
given survey. It can also represent a fraction of the observation name, allowing to list multiple related observations.
The output table offers the option to compute the cross-matches for each observation, with a direct link
to a previous service.

\subsubsection{Body information}
\label{sectionWebBody}

\begin{figure}[h!]
\centering
\includegraphics[width=1.0\textwidth]{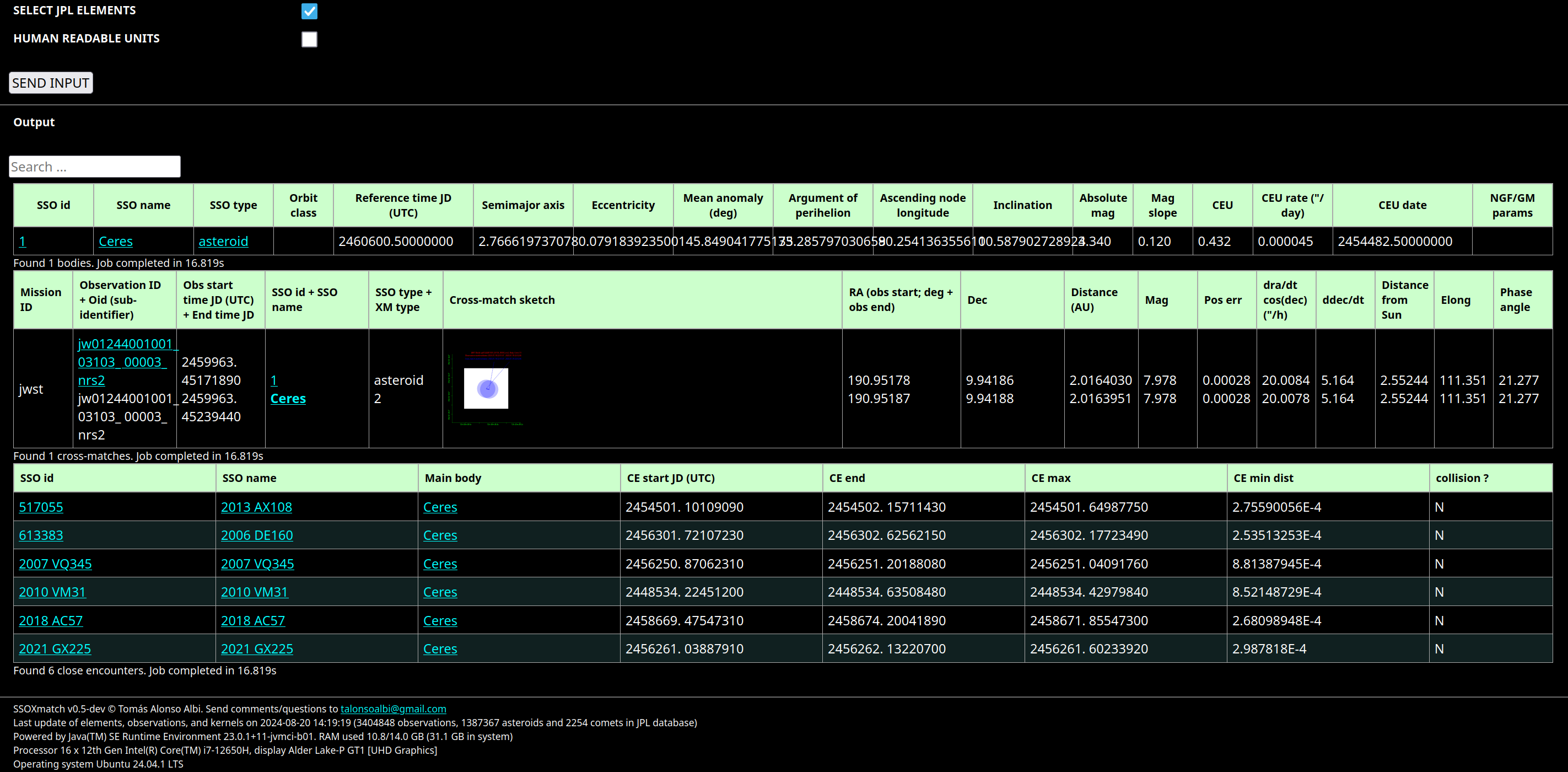}
\caption{Bottom of the page of the service providing information for a particular body.}
\label{figWebBodyInfo}
\end{figure}

This last service is quite interesting, since it provides, for an specific body, the list of
elements present in the database (including the classification of the orbit), the list of cross-matches
in any of the surveys supported (with links to other services in the observation identifiers, to explore
the presence of other bodies as well), and the list of close encounters with other bodies. The screenshot in
Fig \ref{figWebBodyInfo} shows the output for Ceres, showing in this case the close encounter with
it of other asteroids that are considered as massless bodies. In this output the cross-matches
of Ceres have been intentionally limited to one, to show the three tables.

\subsection{Comparison with SkyBoT and JPL services}
\label{sectionWebComparison}

As mentioned previously, the ephemerides service includes an option to compare the output with Horizons and Miriade,
by calling these services and showing their results and the discrepancies with SSOXmatch.
The screenshot in Fig. \ref{figWebEphem} shows a test for the NEO 2015 TC25, from an observatory in
Madrid, Spain, for the Modified Julian Day 57000.0, expressed in UTC. The discrepancy shown
by Horizons is only 0.00001$^{\circ}$, while the Miriade service deviates from the other two by
6.5$^{\circ}$. 
In general, the results of all services agree for most bodies and times, but as explained in previous sections
for bodies that are or were close to the Earth, a considerable deviation may be found in the output
from the LTE services. For very critical bodies
like 2020 CD3, the difference can exceed 50$^{\circ}$, making the ephemerides meaningless. For the Apophis test
described in Sect. \ref{sectionTestApophis}, the output from SSOXmatch for the Julian Day 2463137.5 UTC agree
with Horizons to within 0.00001$^{\circ}$ using JPL elements (or 0.07$^{\circ}$ when using the elements from the
Lowell database), with the body at 1.5 au from the Earth, while in Miriade, using also the elements from the
Lowell observatory, the discrepancy is 2.5$^{\circ}$.

The ephemerides service in SSOXmatch includes a preset option that allows to control some aspects of the numerical
integration to make it more accurate or faster. One of the options provided is to use the kernel of the asteroid
in case the body and input date are within the time window of the kernel. In case of using it, the match of the
position and velocity vector between SSOXmatch and Horizons for the previous test of 2015 TC25 would be perfect.
The kernel will also ensure the little discrepancies will
not grow excesively when the body needs to be propagated more years into the past. As explained in Sect.
\ref{sectionIntegrationSchema}, this approach is more efficient than implementing a better integrator,
that would slowdown dramatically the calculations. As can be deduced from this result and those presented
in table \ref{tableObsResults}, the output for bodies that are not in this situation is also
identical to Horizons, since the numerical integration model is essentially the same. In table \ref{tableObsResults2}
the output of Miriade is shown for comparison with the same table, showing a good agreement, since
these bodies did not suffer any close encounter with Earth.

In SSOXmatch the NEO 2015 TC25 does not require a kernel to reproduce any
of the observations available for it in the Minor Planet Center website, but the situation is different for
2020 CD3. To reproduce the oldest observation for it (May 9, 2018), the kernels for Horizons are needed to get a
consistency around 0.2$^{\circ}$ with the observation. The same result is returned by Horizons. When the kernel
is not used, the discrepancy will be 22$^{\circ}$ and 29$^{\circ}$ for JPL and Lowell elements respectively, while
the discrepancy in Miriade is 120$^{\circ}$. In the most critical situations like this one, the output from
Horizons is the most accurate one available, and without a more rigorous and slow integration model, the
kernels obtained from Horizons for these bodies are required to reproduce the observations with the minimum
possible discrepancy.

\begin{table}[h!]
\footnotesize
\centering
\begin{tabular}{ c c c }
\hline
\textbf{Observation} & \textbf{Miriade} & \textbf{Discrepancy} \\
\textbf{number} & \textbf{(RA, DEC} & \textbf{(")} \\
\hline
1  & 04 26 03.09, -29 29 22.6 & 10.980 \\
2  & 22 55 29.44, -00 30 14.5  & 12.960 \\
3  & 20 39 46.68, -17 45 37.2  & 1.512 \\
4  & 01 38 12.77, -01 41 32.2  & 4.860 \\
5  & 21 03 36.27, -11 07 09.0  & 0.144 \\
6  & 16 38 52.47, 01 25 38.5   & 1.620 \\
7  & 15 42 13.47, -09 41 03.1  & 0.324 \\
8  & 14 32 17.42, -03 03 13.8  & 0.360 \\
9  & 16 06 25.45, -07 38 58.8  & 1.368 \\
10 & 01 21 42.82, 10 05 29.7   & 2.916 \\
11 & 19 59 35.22, -18 46 29.0  & 0.324 \\
12 & 09 35 55.13, -72 19 01.5  & 1.008 \\
13 & 00 01 12.73, 00 08 51.5   & 1.188 \\
14 & 01 21 39.37, 09 33 19.4   & 0.324 \\
15 & 02 34 10.00, 16 42 45.2   & 0.288 \\
16 & 10 44 14.58, -78 40 57.2  & 27.432 \\
\hline
\end{tabular}
\caption{Positions computed with Miriade for the list of observations in table \ref{tableObs}. The
position discrepancy with respect to the observation is shown in arcseconds in the last column.}
\label{tableObsResults2}
\end{table}

\section{Conclusions}
\label{sectionConclusions}

This work presents a new software package aimed to help in the accurate identification of minor bodies
in astronomical images taken in the last decades from any observatory on Earth or in the space. For this
purpose the numerical integration model implemented is consistent with the tools used for the orbit fittings,
allowing to reproduce the astrometry of the observations, usually taken years in the past. 
For bodies suffering close encounters with massive bodies, a set
of kernels were computed with Horizons to keep the maximum possible accuracy or consistency with this server,
allowing to propagate the trajectories for many years without sacrifying performance in the integration process.
The uncertainties in the positions for Lowell and JPL elements are calculated with a common method, transforming the
estimated in-orbit uncertainty for the date to the angular uncertainty as visible from the telescope. Bodies
suffering collisions are correctly handled. The detection of cross-matches follows a direct approach free of
position interpolations, and it is highly optimized. The round-off errors are minimized using the Kahan algorithm
for compensated summation. At the same time, the flexibility allows to support new surveys without any code
change, the performance has been maximized so it is not a problem to apply the software to millions of observations
and minor bodies, and the code quality and documentation (javadoc, JUnit test, and documentation manual) are inline
with the highest standards in software development. The software has been extensively validated using long- and
short-term numerical integration tests, with favorable results in the comparison of the ephemerides and the
cross-matches with other tools. Lastly, a complete set of web services were implemented for an easy testing.

In SSOXmatch there are two databases available to compute cross-matches and ephemerides, that can be selected by
activating or not the corresponding check box in the web service, or by choosing the database in the properties
file when using the software through the command-line. Having at least two databases directly available is
important, since as explained in previous sections the orbits of many bodies have considerable uncertainties.
When a cross-match is found with both sets of elements, the probability to find the body in the image is expected
to be higher, and the identification more reliable.

As described previously, the source code is original and the present work only makes use of public data available in
different NASA and ESA archives. Although the source code is not publicly available, a
\href{http://talonsoalbi.noip.me/ssoxmatch/}{web page} has been created to allow anyone
interested to download and use the executable jar file of the program.

With the flexibility of the implementation offered by the highly customizable input
options (allowing to evolve some aspects of the dynamical model without changing the code), and the performance
achieved in the computations, this tool is ready for the challenges expected in the next years, more specifically
the increment in the number of discovered bodies and total number of observations, arising from the surveys of
the new generation of telescopes.

\footnotesize{
\bigskip
\textit{Acknowledgements}: The author wish to thank the two anonymous referees for their detailed and valuable comments, that helped to improve and clarify many aspects of the paper.
}

\bibliographystyle{elsarticle-harv}


\end{document}